\documentclass[letterpaper,twocolumn,10pt]{article}

\usepackage{usenix,epsfig} 
\usepackage{xspace}
\usepackage[inline, shortlabels]{enumitem}
\usepackage{booktabs}
\usepackage{cprotect}
\usepackage{amsmath}
\usepackage{array}
\usepackage[table,svgnames]{xcolor} 

\usepackage{graphicx}
\usepackage{subcaption}

\usepackage{algorithm}
\usepackage{algpseudocode}
\definecolor{commentgray}{HTML}{696969}

\newcommand{\algrule}[1][.2pt]{\par\vskip.3\baselineskip\hrule height #1\par\vskip.5\baselineskip}

\newcommand{\Ours}{\textsc{CascadeServe}\xspace}

\newcommand{\todo}[1]{{\color{red}TODO: #1}}
\newcommand{\ferdi}[1]{{\color{blue}Ferdi: #1}}
\newcommand{\ziniu}[1]{{\color{purple}Ziniu: #1}}
\newcommand{\lei}[1]{{\color{orange}Lei: #1}}
\newcommand{\nesime}[1]{{\color{violet}Nesime: #1}}
\newcommand{\srm}[1]{{\color{magenta}Sam: #1}}

\renewcommand{\todo}[1]{}
\renewcommand{\ferdi}[1]{}
\renewcommand{\ziniu}[1]{}
\renewcommand{\lei}[1]{}
\renewcommand{\nesime}[1]{}
\renewcommand{\srm}[1]{}

\makeatletter
\patchcmd{\maketitle}
	{\@maketitle}
	{\@maketitle\vspace{-5em}}
	{}
	{}
\makeatother

\begin{document}

\date{}

\title{\vspace{-5em}\Ours: Unlocking Model Cascades for Inference Serving}

\author{
{\rm Ferdi Kossmann$^1$, Ziniu Wu$^1$, Alex Turk$^2$, Nesime Tatbul$^{1,2}$, Lei Cao$^{1,3}$, Samuel Madden$^1$}\\
{$^1$MIT, $^2$Intel, $^3$University of Arizona}\\
{\{ferdik, ziniuw, tatbul, lcao, madden\} @csail.mit.edu, alex.turk@intel.com}
}

\maketitle

\thispagestyle{empty}

\subsection*{Abstract}

Machine learning (ML) models are increasingly deployed to production, calling for efficient inference serving systems. Efficient inference serving is complicated by two challenges: 
\begin{enumerate*}[label=(\roman*)]
    \item ML models incur high computational costs, and
    \item the request arrival rates of practical applications have frequent, high, and sudden variations which make it hard to correctly provision hardware.
\end{enumerate*}
Model cascades are positioned to tackle both of these challenges, as they
\begin{enumerate*}[label=(\roman*)]
    \item save work while maintaining accuracy, and
    \item expose a high-resolution trade-off between work and accuracy, allowing for fine-grained adjustments to request arrival rates.
\end{enumerate*}
Despite their potential, model cascades haven't been used inside an online serving system.
This comes with its own set of challenges, including workload adaption, model replication onto hardware, inference scheduling, request batching, and more.
In this work, we propose \Ours, which automates and optimizes end-to-end inference serving with cascades. 
\Ours operates in an offline and online phase. In the offline phase, the system pre-computes a \emph{gear plan} that specifies how to serve inferences online.
In the online phase, the gear plan allows the system to serve inferences while making near-optimal adaptations to the query load at negligible decision overheads.
We find that \Ours saves 2-3$\times$ in cost across a wide spectrum of the latency-accuracy space when compared to state-of-the-art baselines on different workloads.

\section{Introduction}
The growing capabilities of machine learning (ML) models have prompted a rapid growth in their adoption for a variety of use cases.
For example, Meta performs trillions of model inferences per day, and 70\% of their ML energy consumption is for inference~\cite{meta-scale}. Given the growing scales of ML adoption, it is crucial to build efficient inference serving systems.

\begin{figure}
    \centering
\includegraphics[width=0.48\textwidth]{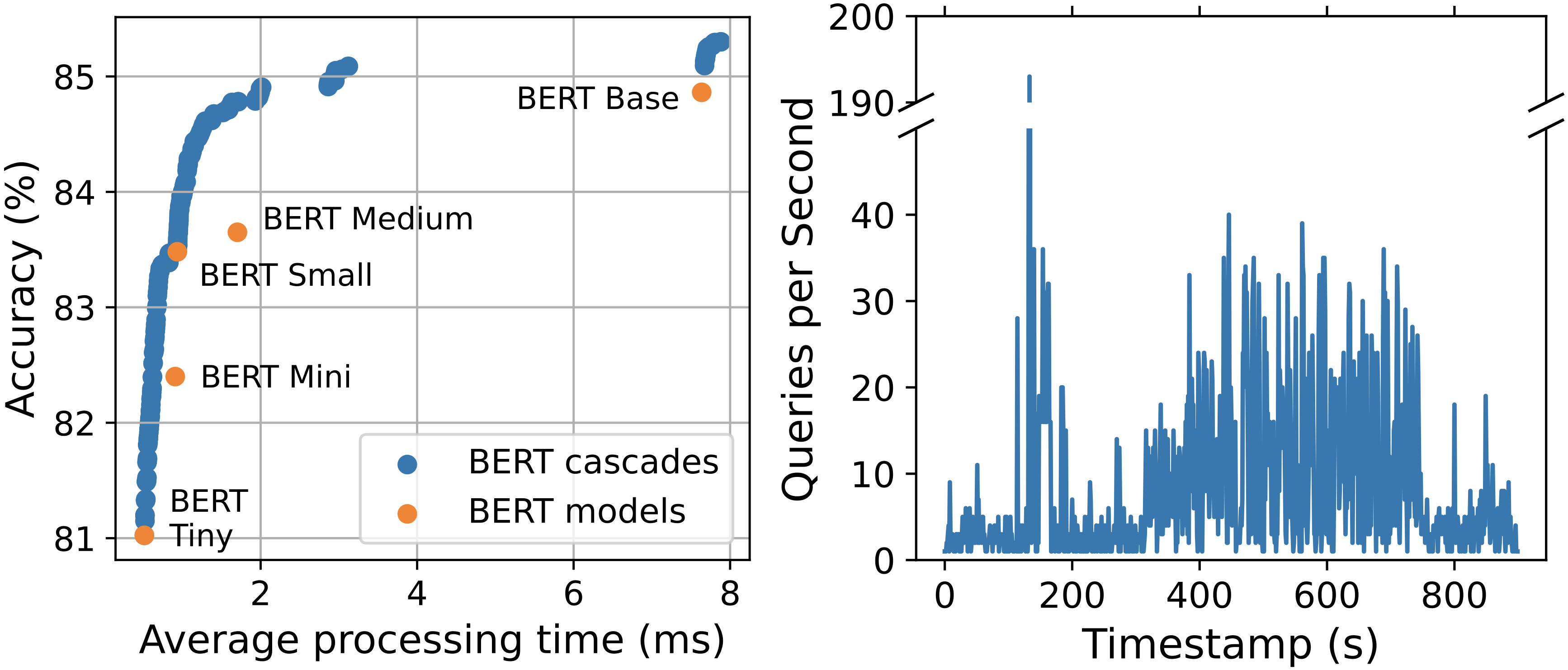}
\caption{Per-sample processing times and accuracy of fine-tuned BERT models on the Sentiment-140 benchmark~\cite{sentiment140}.}

    \label{fig:motivational-intro}
\vspace{-1em}
\end{figure}

Inference serving systems need to optimize their service with regard to three metrics:
\begin{enumerate*}[label=(\roman*)]
    \item they need to serve predictions at low latency,
    \item they need to make predictions with high accuracy, and
    \item they need to incur low cost (i.e., use few hardware resources).
\end{enumerate*}
Achieving satisfactory performance in terms of these metrics is difficult because of two challenges:
\begin{enumerate*}[label=(\roman*)]
    \item ML models incur high computational costs (e.g., Llama-2 70B takes $450ms$ to generate the first token on two V100 GPUs\footnote{With 4-bit GPTQ-quantization~\cite{gptq} and a prompt length of 40 tokens.}), and
    \item the request arrival rate in many applications has frequent, high, and sudden variations, which make it hard to optimally provision hardware~\cite{alpaserve, shepherd}. For example, Figure~\ref{fig:motivational-intro} (right) shows an excerpt of a request trace that is representative of inference serving~\cite{azure-trace}. Provisioning for peak workload leaves hardware resources idle most of the time, but under-provisioning leads to insufficient throughput during workload peaks.
\end{enumerate*}


We find that \emph{model cascades} are positioned to naturally tackle both of these challenges. ML models generally posses a trade-off between runtime and accuracy, where some models are fast but inaccurate (e.g., Llama-7B) while others are slow but accurate (e.g., Llama-70B) ~\cite{scaling-laws1, scaling-laws2}.
Prior work on model cascades~\cite{first-cascade} proposes a way to exploit this trade-off and deliver the accuracy of expensive models with significantly less computation. In a model cascade, each sample is first passed through a cheap model, after which the certainty of the model's prediction is quantified. If the cheap model is certain about its prediction, the prediction is used as the final output. However, if the prediction certainty is below a defined threshold, the sample is forwarded to the next model in the cascade, which is usually more expensive and accurate. 
This allows cascades to predict most inputs with cheap models but predict ``difficult'' inputs with expensive models, simultaneously enabling high accuracy and low computational cost.

Figure~\ref{fig:motivational-intro} shows the accuracy and average processing time of fine-tuned BERT models~\cite{bert} on a sentiment classification task~\cite{sentiment140}. Some model cascades can achieve the same accuracy as the largest model (BERT Base) while taking $3.8\times$ less time to process a sample on average. Furthermore, model cascades expose a high-resolution trade-off between processing time and accuracy. In this work, we leverage both properties to address the challenges imposed by inference serving. 


First, we leverage the savings in processing time to reduce the hardware resources needed for serving. Second, we leverage the high-resolution work-accuracy trade-off to scale throughput by making fine-grained adjustments to the cascade. This avoids the dilemma of provisioning hardware for fluctuating workloads. Specifically, under light load, we use cascades that frequently invoke expensive models and achieve high accuracy; under heavy load, we switch to cascades that use cheaper models to achieve higher throughput. Degrading accuracy during high demand is a technique that is already deployed in production today~\cite{twitter-acc-deg}.

Despite their advantages, model cascades have not been applied inside end-to-end serving systems and doing so incurs a set of unique challenges.
First, the certainty thresholds of the cascade need to be dynamically adapted to changes in system load. For large changes, tuning the thresholds becomes insufficient, and the system needs to remove models from the cascades or add new ones.
To do this, the system has to allocate models between available GPUs, accounting for the fact that loading models can take seconds. 
Moreover, the system needs to schedule inferences as several models in the cascade may contend for compute on the same GPU.
Scheduling decisions must consider \emph{batching} effects, which have a large impact on end-to-end performance: Delaying the inference of a model incurs waiting times but allows more samples to be propagated through the model as one batch, which increases throughput through better hardware utilization.
Overall, optimally using cascades inside a serving system thus involves the optimization of many moving pieces, leading to an immense, exponential search space.


In this work, we propose \Ours to leverage the potential of cascades for inference serving. \Ours efficiently adapts cascades to the incoming workload, which involves making decisions at low overhead while facing an exponential search space. 
To achieve this, the system off-loads most of its decision making to an offline planning phase. During the offline phase, \Ours generates a \emph{gear plan} that specifies how to serve inferences under varying system loads. Given this gear plan, online cascade adaption boils down to switching between gears, allowing \Ours to make near-optimal decisions at negligible overhead. 

\Ours generates gear plans with respect to the user's available hardware and a service-level objective (SLO). The SLO defines a constraint on either accuracy or latency. \Ours will fulfill the latency SLO while optimizing for accuracy or vice versa.
To efficiently choose an optimal gear plan, we decompose the exponential decision space into several sub-decisions and propose a novel algorithm to jointly optimize these sub-decisions. In experiments, our algorithm can find near-optimal gear plans within a few minutes.
In addition, \Ours develops an efficient online serving architecture that operate according to a gear plan. 

We evaluate \Ours on workloads with fast and slow models and find that it consistently outperforms state-of-the-art systems. For most points in the latency-accuracy trade-off space, \Ours incurs 2-3$\times$ less cost than the best baseline in that regime. For a fixed number of GPUs, \Ours significantly beats baselines in terms of both, accuracy and latency. 


In summary, our contributions are as follows:

$\bullet$ We propose an efficient inference serving system that operates in an offline preparation phase and an online serving phase. The system is the first end-to-end serving system that automatically optimizes its service through model cascades.

$\bullet$ We design a novel algorithm that generates a near-optimal \emph{gear plan} in the offline phase. The gear plan describes how to serve inferences in the online phase, considering the user's workload and Service-Level Objective, available hardware resources, and the offered load of the system.


$\bullet$ We evaluate the system's performance on different workload traces and inference tasks.

\section{Background}

In this section, we first provide the background on model cascades, and then review prior work on model inferencing.

\subsection{Model cascades}
\label{sec:overview-def}

Most machine learning tasks inherently pose a trade-off between cost and accuracy~\cite{scaling-laws1, scaling-laws2}. Developers usually train models with different architectures/sizes to explore this trade-off. For example, BERT~\cite{bert}, YOLO~\cite{yolo}, and Llama~\cite{llama-2} models are released in various sizes, where some are faster but less accurate, while others are slower but more accurate. 

Previous work has proposed an elegant method to exploit this trade-off by combining models of different sizes into a \emph{model cascade}~\cite{first-cascade}. First, a model cascade feeds a sample through a cheap model to obtain an initial prediction and an estimated certainty of this prediction. If the cheap model is certain, the prediction will likely be accurate and can be used as the final output. However, if the prediction certainty is below a defined threshold, the sample is forwarded to another model, which is typically more expensive and accurate. This conditional forwarding is repeated until the model prediction is certain or the last model in the cascade is reached.

The key insight behind model cascades is that most samples are ``easy'' to predict (e.g., far away from decision boundaries). For these samples, even a cheap model can deliver accurate predictions. The accuracy gain of more expensive models only comes from performing better on the ``difficult'' samples (e.g., samples that lie close to decision boundaries).
By using cheap models for the ``easy'' inputs and expensive models for the few ``difficult'' inputs, model cascades can achieve high prediction accuracy at low average cost. Therefore, model cascades have been used to optimize various ML tasks, such as classification~\cite{casc-search}, regression~\cite{stage}, object detection~\cite{first-cascade}, recommendation~\cite{casc-use-recommend}, or text generation~\cite{casc-use-llm}.


\subsection{Related work}
\paragraph{Inference serving.} Inference serving systems provide an endpoint that users call to obtain predictions on input samples. We divide prior work on inference serving into two categories: works where users request inferences from a specific model and works where the serving system may choose which model or ensemble of models it uses to serve a request.

Clipper~\cite{clipper}, Rafiki~\cite{rafiki}, HOLMES~\cite{holmes} and Cocktail~\cite{cocktail} serve inferences through carefully designed bagging ensembles. Fundamentally, these systems assume that models in a bagging ensemble can be executed independently in parallel, whereas models in a cascade must execute sequentially. This characteristic of cascades imposes a unique set of challenges and requires different inference serving approaches, which we will explain in details later. INFaaS~\cite{infaas} and Model Switching~\cite{model-switching} maintain a pool of models but do not use them in an ensemble fashion. They use a single model for each request but switch between different models depending on the system load. SuperServe~\cite{superserve} leverages weight-shared SuperNetworks~\cite{supernet} to adapt computational demand to the workload at reduced memory overhead. The Databricks model serving platform~\cite{databricks-multimodel} allows users to split requests between models according to user-defined fractions. 

Other works on inference serving assume that users request inferences from a specific model. Production systems with this functionality include SageMaker~\cite{sagemaker}, TensorFlow Serving~\cite{tfx}, TorchServe~\cite{torchserve}, AzureML~\cite{azureml}, Triton~\cite{triton}, Databricks~\cite{databricks}. There also are academic works that propose serving systems that let users query different models which are hosted on shared hardware~\cite{shepherd, deepplan, alpaserve, reef, irina, pretzel, mark, clockwork}. BATCH allows for batched inference serving on serverless platforms~\cite{batch}. Some works focus on how to share GPUs more efficiently among several tasks for inference serving~\cite{gonzalez-multiplexing, salus, orion, paella, gpulets}. 

In summary, the common difference that these works have to \Ours, is that they don't leverage cascades to optimize inference serving.

\smallskip
\noindent  \textbf{Model cascades.} 
Several works have identified cascading as a way to reduce computational demand while maintaining predictive performance. 
IDK cascades~\cite{idk-cascades} search for cascades on the pareto frontier between compuational cost (i.e. floating point operations) and accuracy. Further works expand on this method~\cite{casc-search, casc-search2, casc-search3}. UnfoldML~\cite{unfoldml} proposes 2-dimensional cascades and optimizes them for spatio-temporal cost and accuracy. Willump~\cite{willump} automatically builds cascades for ML pipelines where the bottleneck is computing the inputs to the model. 
Since these works don’t consider end-to-end serving systems, they don’t address many aspects of leveraging cascades inside such systems (e.g. the allocation of hardware resources, request batching, etc.) To effectively leverage cascades inside a serving system, these factors need to be co-optimized with the cascades themselves.
FrugalML~\cite{frugalml} and FrugalGPT~\cite{frugalgpt} optimize calls to serving endpoints. While they use cascading to optimize calls to APIs, they don't aim to improve efficiency inside a serving system. Finally, many works have proposed the use of specific model cascades to optimize a particular prediction task. Noscope~\cite{noscope} and WEG~\cite{weg} use cascades to accelerate searching through large videos to find objects. Stage~\cite{stage} uses cascades for runtime estimation inside a database engine. 


In summary, the common differences between these works and \Ours is, that these works don't consider an end-to-end serving scenario and the aspects that are involved to leverage cascades in this context.

\smallskip
\noindent  \textbf{ML pipeline serving.} Some works seek to optimize the inference of entire ML pipelines, where several different models may interact with each other to produce outputs (e.g. using Alexa might first involve speech recognition, then speech-to-text transcription, then answering the question, then text-to-speech generation). InferLine~\cite{inferline} automatically scales each stage in such a pipeline, assigns stages to appropriate hardware and tunes model batch sizes. Nexus~\cite{nexus} optimizes video processing pipelines by assigning models to GPUs and adapting batch sizes for parts of a model (instead of models as a whole). Skyscraper~\cite{vetl} optimizes video processing pipelines through buffering and cloud bursting techniques. The most fundamental and common difference between these works and \Ours is that these works optimize a user-specified ML pipeline whereas \Ours serves inferences by building cascades. 






\section{System overview}
\label{sec:overview}

In the previous sections, we described how leveraging the potential of cascades in an inference-serving setting poses a unique set of challenges that prior work does not address. We now provide a high-level overview on how \Ours tackles these challenges.


\smallskip
\noindent  \textbf{User-provided information.} 
In \Ours, users register their workload by providing 
\begin{enumerate*}[label=(\roman*)]
    \item a set of trained models for building cascades (e.g., Llama2-7b, Llama2-13b, Llama2-70b~\cite{llama-2}),
    \item a set of labeled samples for evaluating the performance of the model cascades (e.g., a public LLM benchmark~\cite{hellaswag}), and
    \item the users' available hardware resources (e.g., number of GPUs, memory).
\end{enumerate*}
Additionally, users can tailor \Ours's service to their needs by specifying a Service-Level Objective (SLO). For latency-sensitive applications (e.g., recommender systems), users may set an SLO on latency, instructing \Ours to optimize accuracy without surpassing the latency target. Analogously, for accuracy-sensitive applications (e.g., code generation models), users may set an accuracy SLO, instucting \Ours to optimize latency while achieving the desired accuracy.

\begin{figure}
    \centering
    \includegraphics[width=0.47\textwidth]{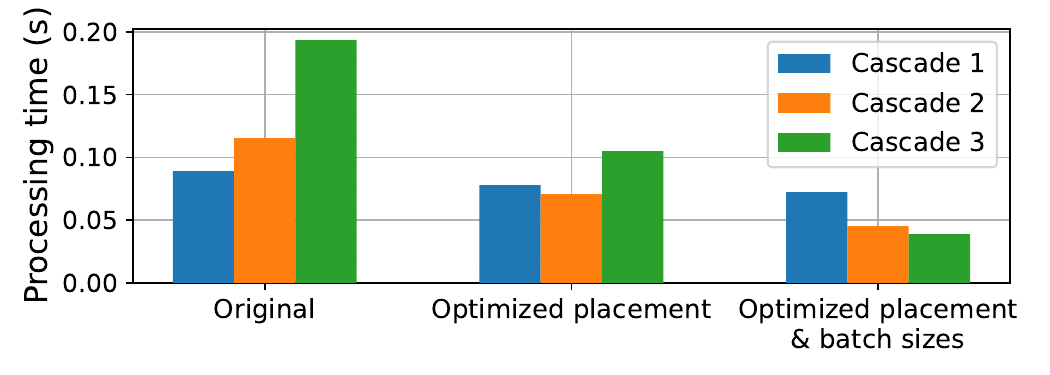}
    \caption{Processing times of BERT cascades when changing model placement and batch sizes.}
    \label{fig:interdependence}
    \vspace{-1em}
\end{figure}

\begin{figure*}
    \vspace{-2.5em}
    \centering
    \includegraphics[width=0.8\textwidth]{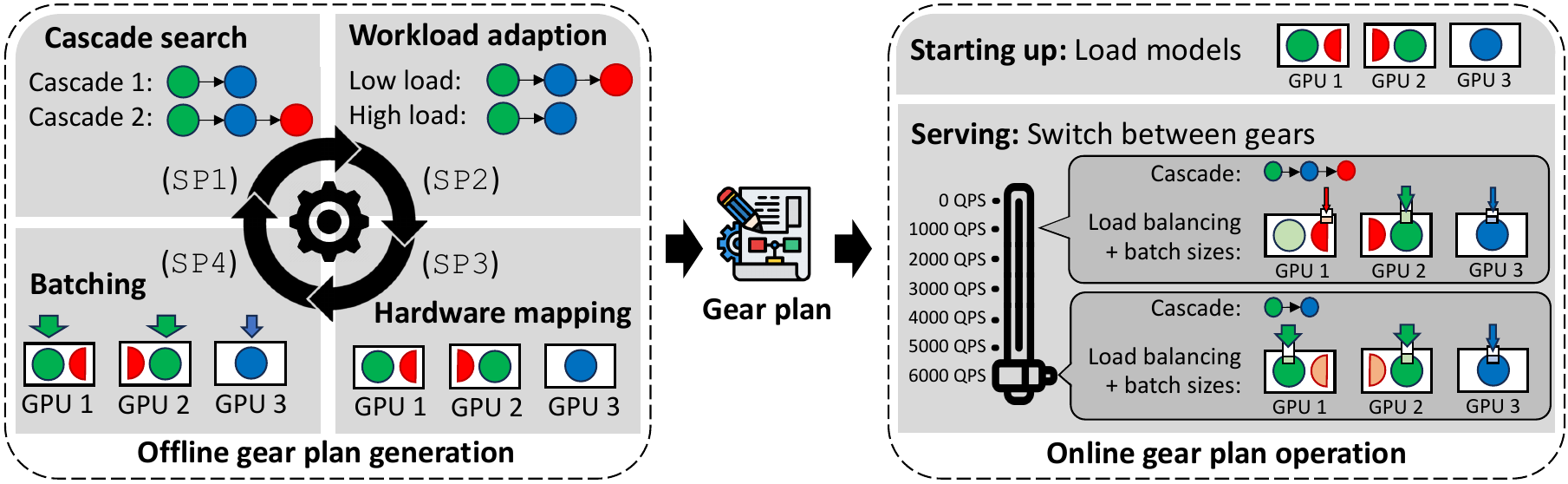}
    \caption{Overview of \Ours.} 
    \label{fig:overview}
\end{figure*}

\smallskip
\noindent  \textbf{\Ours workflow.}
Given the user's inputs, \Ours optimizes its service by adaptively switching between cascades.
Figure~\ref{fig:interdependence} shows that different cascades can incur significantly different processing times (i.e., latency). 
\Ours adapts its throughput by switching to fast cascades whenever the system load is high. For example, given the system configuration on the left (``original''), \Ours may use Cascade 1 to meet the required throughput whenever the system load is high. However, when the system load is low, a more accurate cascade may also fulfill the latency SLO, even if it is slower than Cascade 1.

The switching task is not straightforward because the latency/throughput depends not only on the cascade. To cope with high query loads, models within the cascade need to be replicated across GPUs. Model replication is not a trivial task since every GPU has a finite amount of memory (VRAM) that generally can not fit all models. Figure~\ref{fig:interdependence} shows how, after optimizing model placement, the processing times of cascades can dramatically change.
Therefore, cascades must be tuned jointly with the placement of models on the hardware. 
Furthermore, in model serving systems, it is common to batch samples together and forward them through the ML model as one tensor. Larger batch sizes can significantly improve throughput by improving the utilization of the underlying hardware~\cite{clipper}. However, larger batch sizes also incur waiting time since inference is delayed until the batch is filled. Figure~\ref{fig:interdependence} shows how, after optimizing the batch sizes, Cascade 3 becomes the fastest cascade despite originally being the slowest. Therefore, \Ours must also consider and optimize for batching effects.

In summary, \Ours must optimize and carefully coordinate many interdependent moving pieces, which impose an exponential search space of possible options.
Moreover, this optimization must incur a low overhead on the critical path of cascade inference.

\Ours tackles this problem by offloading most of the decision-making to an offline planning phase before serving users' queries online. Specifically, \Ours generates a \emph{gear plan} that determines how to serve inferences online when faced with different query loads. The gear plan further determines how models should be replicated across GPUs. \Ours uses a fixed allocation of models to GPUs throughout online serving, which allows it to quickly switch between cascades without having to wait for models to be loaded. Figure~\ref{fig:overview} summarizes the life cycle of \Ours: First, a gear plan is generated offline. Then, as the system is started, all models are loaded onto GPUs as specified by the gear plan. Finally, \Ours performs inferencing by simply switching between gears, allowing it to make near-optimal decisions at negligible runtime overhead. 

\smallskip
\noindent  \textbf{Gear plan generation.} 
\Ours generates a gear plan by decomposing the decision problem into four subproblems.
First, the system must construct cascades from the given set of models (\verb|SP1|). Different cascades vary in accuracy and throughput, and we want to identify a subset of cascades on the Pareto-optimal frontier.
Second, the system must decide which cascades to use for each QPS range (\verb|SP2|). Higher QPSes require cascades with higher throughput. 
Third, cascades must be distributed and replicated across the available hardware resources (\verb|SP3|). High GPU utilization can only be achieved if different models are collocated onto the same GPU. However, GPU VRAM is a scarce resource, and different cascades must co-exist in VRAM.
Fourth, the system must select a batch size for each model replica (\verb|SP4|). Large batch sizes allow for improved throughput but incur higher latency since inference is delayed until a batch is filled.


Each of the four problems has an exponential search space. Furthermore, the subproblems are interdependent and, therefore, have to be optimized jointly. 
To do so, \Ours designs a novel Expectation-Maximization-based algorithm~\cite{expectation-maximization}. For each subproblem, \Ours introduces a submodule that optimizes the subproblem with respect to a fixed solution to all other subproblems. \Ours then alternately calls these submodules to derive a joint solution. During this process, the joint solution is iteratively improved by optimizing one subproblem at a time. Our algorithmic design of the submodules and iteration logic leads to a fast convergence, which we prove in Appendix~\ref{sec:convergence}. We give a detailed description of the algorithm in Section~\ref{sec:offline}. 


\smallskip
\noindent  \textbf{Gear plan operation.} \Ours uses a Producer-Consumer architecture~\cite{producer-consumer} to serve inferences online. The producer measures the QPS by periodically counting the number of queries arriving over a fixed interval (e.g., 100ms).
Then, it can quickly look up the optimal gear from the gear plan according to the QPS.
The producer forwards requests to \emph{inference server}, which contains a queue for each model in the cascade that buffers requests before inference. 
The producer places the requests into the queue of the first model in the gear's cascade.
Each queue is periodically polled by the \emph{consumer}, which triggers inference if enough samples are queued according to the batch size in the current gear.
This design is highly scalable because producers, consumers, and inference servers can be scaled independently to avoid bottlenecks. We give a detailed description of this architecture in Section~\ref{sec:online}.

\section{Gear plan generation}
\label{sec:offline}

The gear plan dictates how requests are served online. Specifically, the gear plan specifies the placement of models on users' hardware devices and assigns a gear to each QPS range that indicates the model cascade and batch size to use for this range.
Then, during online serving, \Ours measures the incoming QPS and switches between gears accordingly. 

In the following, we describe how \Ours decomposes the complicated problem of searching for optimal gear plans into four simpler subproblems. Solving each subproblem individually is much more tractable than solving the original problem.
More importantly, based on this  \Ours  
, and the decomposition still allows us to jointly optimize the interdependent subproblems, which \Ours achieves via an iterative algorithm that is inspired by Expectation-Maximization~\cite{expectation-maximization} (EM). We describe this algorithm in Section~\ref{sec:opt-coop}. We then discuss the submodules that optimize the individual subproblems. By leveraging insights specific to cascades, each submodule is capable of finding a near-optimal solution to its subproblem.


\subsection{Subproblem co-optimization}
\label{sec:opt-coop}

\begin{algorithm}[t]
\small
\caption{Gear plan optimization}
\label{alg:offline-algo}
\begin{algorithmic}[1]
\vspace{.2em}
\Statex \hspace{-14pt} \textbf{Inputs:} SLO, model\_profiles, qps\_distribution
\algrule

\Statex \textcolor{commentgray}{// Initial gear plan}
\State gear\_plan $\gets$ init\_plan(qps\_distribution, model\_profiles, SLO)
\State error\_code $\gets$ "ok" 
\State subproblem\_modules $\gets$ [search\_cascades, assign\_cascades, place\_models, tune\_batch\_sizes]
\State cur\_subproblem $\gets 0$

\vspace{.5em}
\Statex \textcolor{commentgray}{// Optimize one subproblem at a time keeping the others fixed}
\While{not converged}
    \If{cur\_subproblem == -1}
        \State \textbf{raise} Error("infeasible")
    \EndIf

    \State module $\gets$ subproblem\_modules[cur\_subproblem]
    \State error\_code, gear\_plan $\gets$ module(error\_code, gear\_plan)

    \If{error\_code == "ok"}
    \Statex \textcolor{commentgray}{  \ \ \ \ \ \ \ \ \ \ \ \  //Go to the next submodule and optimize}
        \State cur\_subproblem $\gets$ (cur\_subproblem + 1) \% 4
    \Else
    \Statex \textcolor{commentgray}{ \ \ \ \ \ \ \ \ \ \ \ \ //Go to the previous submodule and resolve the error}
        \State cur\_subproblem $\gets$ cur\_subproblem - 1
    \EndIf
\EndWhile
\end{algorithmic}
\end{algorithm}


As discussed previously, \Ours employs a submodule for each optimization subproblem. These submodules are iteratively optimized with respect to a fixed solution of the other subproblems. \Ours cycles through the submodules to derive a joint solution at convergence. Algorithm~\ref{alg:offline-algo} shows pseudo-code for this procedure, and we put the proof of its convergence in Appendix~\ref{sec:convergence}.

\paragraph{Iteration logic.} In line 3 of Algorithm~\ref{alg:offline-algo}, \Ours cycles through subproblems in the following order: 
\begin{enumerate*}[label=(\roman*)]
        \item searching for cascades,
        \item assigning cascades to each QPS range,
        \item placing models onto hardware (including load balancing), and
        \item assigning a batch size of each model replica for each QPS range.
\end{enumerate*}
We create one submodule to optimize one subproblem.
Submodules throw an ``infeasible'' error when they cannot find a \emph{feasible} gear plan, i.e., a gear plan that fulfills the SLO and ensures the resource capacities (e.g., the number of GPUs and memory) are not exceeded.

The motivation behind \Ours's error-driven approach is that submodules need to make choices with respect to a fixed solution of all other subproblems. In \Ours, the submodules make optimistic decisions and assume that, even if their decisions might not produce a feasible gear plan, the other submodules can optimize their solutions to make the plan feasible. For example, when the workload adaption module (\verb|SP2|) assigns cascades to QPS ranges, it needs to know how much throughput each cascade achieves, which may be further improved by the model placement module (\verb|SP3|). Therefore, instead of missing out on an effective cascade that may become feasible after the optimizations of other submodules, the current submodule needs to consider currently infeasible cascades. An error will be thrown if the other submodules can not optimize the cascade further.
In this case, the current submodule will backtrack this error and correct its previous solution. This error-driven approach ensures that each submodule will not miss the effective gear plan.

When a submodule throws an error (lines 13-14 in Algorithm~\ref{alg:offline-algo}), it is caught by the previous submodule in the sequence, which will try to adjust its solution based on the error code. If this is not possible, this submodule throws an error to the one before. This error resolution process continues recursively. If the error can't be resolved at the first submodule, a final error will be raised to the user (lines 6-7), e.g., "impossible to meet the SLO given the provided hardware resource."
Once all errors are resolved, \Ours keeps iterating until it converges (lines 11-12).
In the following sections, we provide details on how each module optimizes the corresponding sub-problem, detects an error, and resolves it. 



Our EM algorithm requires an initialization (line 1) on model placement (\verb|SP3|) and batch sizes (\verb|SP4|) to start the optimization of searching for cascades (\verb|SP1|) and assigning cascades to QPS ranges (\verb|SP2|). 
We trivially initialize the model placement so that each model is replicated across all GPUs, which may not be feasible.
We initialize the batch sizes of all models to the minimum batch size of 1. This initialization generally applies to any workload, and we empirically verify that our algorithm performs well with this initialization.




\subsection{Cascade search}
\label{sec:opt-casc}

Different cascades achieve different throughput and accuracy and only a subset of cascades is Pareto-optimal. The models and certainty thresholds used in a cascade directly determine its throughput and accuracy. In addition,  the placement and batch size of each model also affect the throughput. 

\Ours forms cascades from a family of models that the user registers (e.g., a set of Llama models or a set of BERT models).  Each model is assumed to produce a {\it certainty}, indicating its prediction confidence. For example, for a model like Llama that outputs a probability distribution on the next possible token, the prediction certainty can be measured as the softmax of the predicted logits. 
We discretize the continuous range of possible certainty thresholds into selectable thresholds.
In practical settings, the number of registered models is relatively small (e.g., there are 3 Llama-2 variants) and a small number of discretized thresholds is sufficient to ensure cascade diversity. 
This allows \Ours to use a simple approach for this submodule: we randomly sample cascades and thresholds and retain only the Pareto-optimal ones.
To evaluate the throughput and accuracy of a cascades, \Ours uses a simulator. Compared to measuring the performance on a real workload, the simulator can evaluate a cascade at low cost. This allows the submodule to sample a large set of cascade, closely approximating the Pareto frontier. 

It is a known result that inference serving systems can accurately be simulated~\cite{clockwork, alpaserve}. We describe and evaluate our simulator in Appendix~\ref{sec:simulator}.

\smallskip
\noindent  \textbf{Error handling.} It is worth noticing that this submodule always includes the cheapest cascade and the most accurate cascade. Receiving an error suggests that the other submodules have failed to attain the SLO, even when always using the cheapest cascade (in case of a latency SLO) or the most accurate cascade (in case of an accuracy SLO). In this case, the user's SLO is unattainable on the user's hardware and the submodule raises an error to the user.

\subsection{Workload adaption}
\label{sec:opt-qps}

%

The output of the cascade search submodule (\S\ref{sec:opt-casc}) is a set of cascades with their accuracies and throughputs. We now describe the submodule that decides which of these cascades to use under different query loads. Specifically, the user specifies the maximum QPS $qps_{max}$ that the system may encounter. This submodule then divides the spectrum of possible QPS from 0 to $qps_{max}$ into $n_{ranges}$ equal-sized QPS ranges and assigns a cascade to each range. 

Intuitively, the submodule could simply assign cascades based on their accuracy and throughput as indicated by the cascade search submodule (\verb|SP1|). However, while the accuracy of a cascade won't change, its throughput might change significantly after other submodules optimize their subproblems to better support the chosen cascades. This naive approach therefore precludes the system from choosing cascades that have insufficient throughput with respect to the current placement and batch sizes but would achieve it after running the other submodules. 

Instead, this submodule starts out by assigning the most performant cascade according to the non-SLO metric to each QPS range (i.e., the most accurate cascade if the SLO is on latency, or the cheapest cascade if the SLO is on accuracy). The submodule only corrects its choices once the other submodules indicate that they cannot fulfill the SLO given the current cascade assignment. Upon receiving an error, the submodule \emph{downgrades} the cascade at the QPS range that the error specifies (e.g., if the user has set a latency SLO and this submodule receives an error that the latency SLO cannot be fulfilled for the QPS range of 100-200, the submodule assigns the next cheaper cascade to that QPS range).


If this submodule receives an \verb|ok| code, it has already found a feasible assignment but might be able to improve it. Specifically, the cascade search module (\verb|SP1|) may have output new cascades that allow for even better cascade assignments. For each QPS range, this submodule iterates through the new cascades and only swaps a new candidate cascade for the current one if the new cascade is better or equal in terms of both, accuracy and throughput.

\subsection{Hardware mapping}
\label{sec:opt-placement}

We now describe the submodule that maps the workload to the underlying hardware. This involves replicating the models across devices (\emph{model placement}) and determining what fraction of the workload each replica should serve (\emph{load balancing}). The goal of this submodule is that for any QPS range, the assigned cascade can be run at the required throughput without loading models into GPU VRAM at runtime. 

Note that we are placing models, not cascades, and different models inside a cascade might need to be placed onto different GPUs. 
Recall that the initial model placement is to replicate each model on every GPU. Although this placement allows for maximum throughput, it generally will not fit into the GPUs' VRAM.
Thus, the submodule designs a greedy pruning strategy that prunes the model placement until it doesn't exceed the memory capacity of any GPU. A final set of models is selected based on their effectiveness at balancing load among GPUs.

\smallskip
\noindent  \textbf{Load balancing.} This module finds the optimal assignment of QPS $q_r$ to model replicas $r$ given the cascade used to serve the load, the placement of models onto hardware, and the QPS $QPS_m$ that each model $m$ in the cascade needs to serve.\footnote{$QPS_m$ is determined through the fraction of samples that the cascade forwards to each model on a validation set, multiplied by the total QPS.}
It assigns load to replicas using the following linear program with decision variables $q_r$.

\begin{alignat}{5}
  & \text{minimize }   & \quad & \, \sum_{r\in \mathcal{R}}q_r                         &       & \textrm{with } q_r \geq 0              & \label{eq:linpro1} \\
  & \text{subject to } & \quad & \sum_{r \in \mathcal{R}[m]} q_r \geq QPS_m            &       & \forall m\in \textrm{cascade}    & \label{eq:linpro2} \\
  &                    & \quad & \sum_{r \in \mathcal{R}[d]} q_r*runtime(r) \leq u     & \quad & \forall d\in \textrm{devices}   & \label{eq:linpro3}
\end{alignat}

where $\mathcal{R}$ is the set of model replicas, and $\mathcal{R}[m]$ denote the replicas of model $m$ and $\mathcal{R}[d]$ denote the replicas on device $d$. 
$runtime(r)$ is the runtime per sample of replica $r\in\mathcal{R}$ with a default batch size of 1. 
Equation~\ref{eq:linpro1} ensures that the load balancer doesn't assign more load than needed. Equation~\ref{eq:linpro2} ensures that that the replicas jointly serve the $QPS_m$ that is demanded by each model $m$ in the cascade.
$u$ is a threshold on the maximum allowed GPU utilization. Equation~\ref{eq:linpro3} enforces that each GPU must have a utilization below $u$.
\srm{qps x runtime does not seem to result in a percentage or fraction, so this equation doesn't seem to be in the correct units.} The load balancer tries to find a load assignment that minimizes $u$ by first building a linear program with $u=100\%$. If the program is solvable, the load balancer builds further linear programs where $u$ is iteratively decreased until $u$ is so low that the program becomes unsolvable. The load balancer returns the minimum value of $u$ for which the program was solvable --- this indicates the lowest GPU utilization that the given model placement allows for. If the program is unsolvable for $u=100\%$, this is indicated through an error.
\todo{This module and the BS error: XPUT or SLO? Should they be handled differently?}
\srm{Does this process ensure that each cascade is fully loaded onto a single GPU?  Or can some cascades span multiple GPUs?  Where is the cost of transmitting data between GPUs accounted for?}

\smallskip
\noindent  \textbf{Model placement.} This submodule greedily prunes model replicas from devices until the replicas assigned to each device fit within that device's memory. 
The utility of pruning a replica $r$ is determined by two factors. First, how much overallocated memory is freed  by pruning it. 
Second, the replica's importance for load balancing. The importance of a replica for load balancing is determined by calling the load balancer for every QPS with a cascade that uses replica $r$. The load balancer is called with a placement where $r$ is pruned and the maximum GPU utilization $u_{max}(r)$ over all QPS is taken. If $r$ is the last replica of a model (i.e., pruning $r$ makes a cascade unrunnable) or the load balancer throws an error, the utility $util(r)$ of pruning $r$ is set to $-\infty$. 

Equation~\ref{eq:prune-util} shows the combined utility for pruning replica $r$, where $m_{over}(d)$ is the overallocated memory on device $d$ and $m_{freed}(r,d)$ is the memory that would be freed on device $d$ when pruning replica $r$. The submodule iteratively prunes the replica with the highest utility until no device's memory capacity is exceeded. If no pruning utility is greater than zero, no placement can realize the given cascade, and this submodule throws an error.

\begin{equation}
util(r) = \frac{\sum_{d\in\textrm{devices}}\max(\,0, m_{over}(d) - m_{freed}(d,r)\,)}{u_{max}(r)}
\label{eq:prune-util}
\end{equation}

\smallskip
\noindent  \textbf{Error handling.}
This submodule may receive an error from the batch size optimization submodule (\verb|SP4|), if the required throughput cannot be achieved for a given QPS range. The error message will indicate the QPS range and the first model $m$ (e.g., Llama-70b) in the cascade where throughput was insufficient (i.e., $m$ is the bottleneck). To resolve the error, this submodule will try to improve $m$'s throughput by including an additional replica of $m$ in the model placement. This can be enforced by setting the utility of pruning $m$ to $-\infty$ for all replicas of $m$, once pruning a replica would violate the additional constraint. If no placement can fulfill this constraint, this submodule also throws an error.

\subsection{Dynamic batching}
\label{sec:opt-bs}

We now describe how \Ours optimizes request batching. Larger batch sizes increase throughput but incur waiting times since samples wait in a queue until the batch is filled and inference is triggered. The optimal batch size depends on the QPS that are assigned to a model replica: At low QPS, it takes long to form large batches as model queues fill slowly. However, the model replica also needs to achieve lower throughput, making small batch sizes ideal. Analogously, high QPS mean that queues are fill quickly and more throughput is required, making large batch sizes suitable.
To account of this, \Ours dynamically adapts the batch size of each model replica to its incoming QPS.

During online serving, samples are buffered for inference in a queue. \Ours triggers inference once a \emph{minimum queue length} is reached. The value of this minimum queue length is optimized by this submodule. The optimal value is determined with respect to a QPS range and is stored in a gear. 

For every QPS range, this submodule starts by setting the minimum queue length to 1 for all replicas of the first model in the cascade. The cascade will then be simulated using \Ours's simulator (\S\ref{sec:simulator}).
When throughput is found to be insufficient, the submodule will increase the minimum queue length of the first mode in the cascade by 1. Since this means that larger batches will be formed, and more samples will be cascaded per batch, increasing the minimum queue length of the first model in the cascade also increases how many samples are included in each batch of the later models.
This process repeats until the required throughput is met. 
The submodule will throw an error if the required throughput cannot be achieved by increasing the batch size (e.g., the wait time will be too long, or the large batch size will exceed memory). 

Prior work~\cite{clipper} proposes a dynamic batching approach tailored to inference serving for a single model or bagging ensembles, which uses a \emph{maximum batch size}. However, this approach induces large latencies in the setting of model cascades because it results in a high resource contention among different models in a cascade.

\section{Gear plan operation}
\label{sec:online}

\begin{figure}
    \centering
    \includegraphics[width=0.485\textwidth]{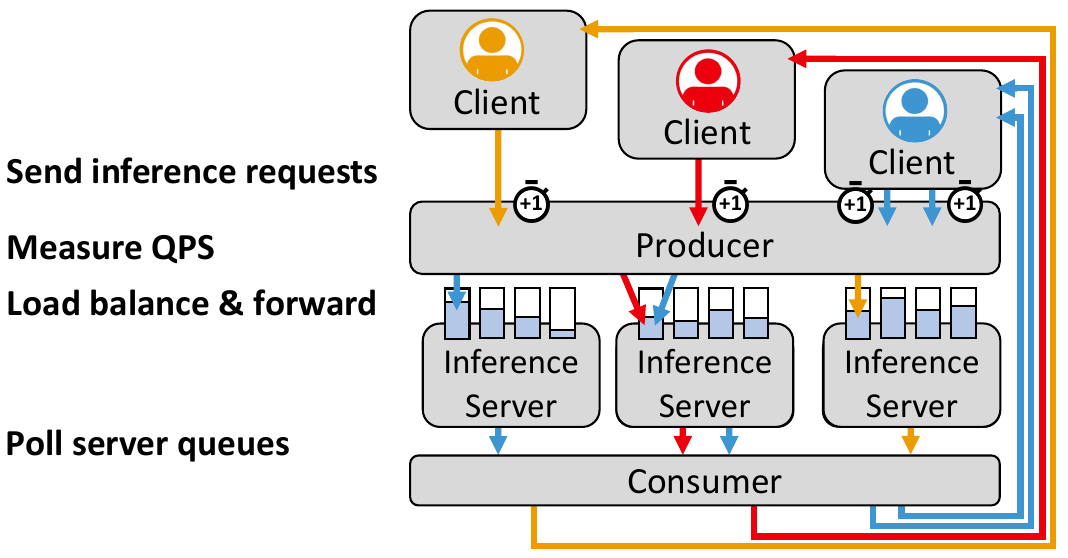}
    \caption{Online serving architecture.}
    \label{fig:serve-architecture}
\end{figure}

We now describe how \Ours serves predictions online. We implemented a serving system that operates according to a gear plan. Given the gear plan, online serving becomes trivial, which allows for performant, scalabale and easily maintainable implementations. Our architecture follows the Producer-Consumer paradigm~\cite{producer-consumer} and is depicted in
Figure~\ref{fig:serve-architecture}, where each grey box depicts a separate process. This architecture is highly scalable as there is no centralized component and consumers, producers, and inference servers can be scaled independently.

\paragraph{Inference server.} An inference server hosts and runs the models. It manages a set of GPUs and loads the models accoring to the gear plan. For a maximum casade length of $\ell_{max}$, the server allocates $\ell_{max}$ memory regions in GPU VRAM to serve as model queues.
An inference server may manage more than one GPU, for example, because a model may not fit into the memory of one GPU. In this case, our implementation runs each model in its own process --- this allows models on disjoint sets of GPUs to run in parallel. 

\smallskip
\noindent  \textbf{Producer.} The producer takes user requests and puts them into the queue of the first model in the cascade. The producer decides which inference server to use based on the load balancing defined in the gear plan. The producer measures QPS periodically by counting the number of queries within the measurement period. 
Based on the measured QPS, the producer switches the gears of the inference servers.

When there is high QPS followed by low QPS, \Ours must avoid downgrading the gear while there are still many samples queued from the high QPS interval. When switching gears on an inference server, \Ours therefore also considers the queue length $Q_0$ of the first model on that server. Specifically, given the measured QPS $qps$, the gear is not changed if $qps < \alpha*Q_0$, where $\alpha$ is a tunable parameter. We find that $\alpha=8$ works well for most scenarios.

\smallskip
\noindent  \textbf{Consumer.} The consumer periodically polls the queues on the inference server and triggers inference based on the batching mechanism described in \S\ref{sec:opt-bs}. 


\section{Evaluation}

In this section, we thoroughly evaluate \Ours to answer the following questions:

\begin{enumerate}
    \item[\S\ref{sec:eval-performance}]  How does \Ours compare against baselines in terms of cost, latency and accuracy?
    \item[\S\ref{sec:eval-adaptivity}]  How does \Ours degrade performance based on the SLO?
    \item[\S\ref{sec:eval-planner}]  How effective is \Ours's gear plan optimization method?
    \item[\S\ref{sec:eval-ablation}]  How much do individual optimizations contribute to \Ours overall performance?
\end{enumerate}

\subsection{Workloads}
\label{sec:eval-workloads}


A workload in our evaluation consists of 
\begin{enumerate*}[label=(\roman*)]
    \item a set of models,
    \item a benchmark task to measure the predictive performance of the cascade, and
    \item a workload trace to emulate when queries are issued against the system.
\end{enumerate*}
We evaluate \Ours on two workloads:  one comprising a family of fast models (BERT) and one comprising a family of slow models (Llama).

\smallskip
\noindent  \textbf{BERT.} We evaluate BERT~\cite{bert} on the commonly used Sentiment-140 benchmark~\cite{sentiment140}, which performs sentiment analysis on Tweets. 
We use BERT Tiny, BERT Mini, BERT Small, BERT Medium, and BERT Base models~\cite{bert-variants} to construct different model cascades.
We use the pre-trained weights of these models and fine-tune them on a training set from this benchmark.
We evaluate the performance of \Ours and other baselines by prompting the systems with samples from a held-out test set of the benchmark.
Each system uses PyTorch~\cite{pytorch} and the Huggingface Transformers package~\cite{huggingface} to invoke the models.

The request pattern is derived from the time stamps at which the Tweets were posted, as done in prior works~\cite{infaas}. We remove all seconds with 0 QPS and take the first 20 minutes of the resulting trace for the experiments. To sufficiently stress the systems, we linearly scale the QPS for each second such that the maximum is 7600 QPS. For the experiment with the accuracy SLO of 81.5\% in Figure~\ref{fig:perf-bert}, we scale the trace such that the maximum is 80,000 QPS. Without the scaling, the workloads would be trivial to run and performance differences in the systems would not be visible. The experiment with the low accuracy SLO needs to be scaled further since the systems use smaller models, which are sufficient to fulfill the SLO.

\smallskip
\noindent  \textbf{Llama.} We use Llama-2 7b, Llama-2 13b, Llama-2 70b~\cite{llama-2} as well as OpenLlama 3b~\cite{openllama}. For all models, we consider a 4-bit GPTQ-quantized version. We use ExLlama~\cite{exllama} to invoke the models. 
We evaluate the Llama models on the commonly used HellaSwag~\cite{hellaswag} benchmark. 

We derive the workload pattern from an invocation trace of Microsoft Azure Functions~\cite{azure-trace}, which is commonly used in other works as well~\cite{alpaserve, shepherd}. \srm{Say this is shown in Figure 1 (right?)} We randomly sample a 20 minute window from the trace and linearly scale the trace so it has a maximum of 60 QPS, and a maximum of 196 QPS for the experiment in Figure~\ref{fig:perf-llama} with an accurcy SLO of 55\%.
Like for the BERT workload, the scaling is necessary to make it feasible to run the workload on our hardware resources while still sufficiently stressing the systems and making the workload non-trivial to run.

\subsection{Experimental set up}
\label{sec:eval-setup}

\paragraph{Hardware.} We run the experiments on NVIDIA Tesla V-100 GPUs with 32 GB of VRAM. Our cluster provides limits us to a  maximum of 16 concurrent GPUs , which is the reason for the workload scaling described in~\ref{sec:eval-workloads}. The GPUs are distributed across 8 nodes, where each node has 2 GPUs that are connected over a PCIe bus.

\smallskip
\noindent \textbf{Implementation.} We implemented \Ours on top of Ray~\cite{ray}. The producer and consumer are Ray actors and run in their own process.  Inference servers are actors too and each inference server manages a set of GPUs exclusively. Requests are issued from a separate client process that runs in an open loop (i.e. issues requests without waiting for the result of previous requests).

In all experiments, \Ours estimates the certainty of the models using the output logits (\S\ref{sec:certainty}). However, the certainty estimation method is not fundamental to our implementation and can easily be exchanged.

\smallskip
\noindent  \textbf{Baselines.} We compare against three baselines.

First, we use a Dynamic Batching (DynBa) baseline. DynBa statically provisions GPUs and uses the same model for all inferences. It uses dynamic batching to optimize throughput and latency. We use a similar batching mechanism as used in \Ours because it delivered the best results when compared to alternatives. 


Second, we compare against Cocktail~\cite{cocktail}. Cocktail is a state-of-the-art academic system that uses bagging ensembles and autoscaling. 
Cocktail's open-source implementation is tightly coupled to AWS, and we re-implemented it to run in our setup. Furthermore, we make the following improvements and denote the resulting system as \emph{Cocktail+}: 
\begin{enumerate*}[label=(\roman*)]
    \item Cocktail uses a forecasting model for autoscaling. Cocktail+ is provided with the groundtruth workload forecast to eliminate any error that the forecasting model might introduce.
    \item VMs may take dozens of seconds to boot before they are available. When Cocktail+ requests additional VMs, they are immediately available. After requesting the VM, Cocktail+ can immediately load the desired model into VRAM. We then perform one warmup inference in the background before samples are routed to the newly spawned instance because we found that this warmup greatly improves performance.
    \item We implement dynamic batching for Cocktail+ using \Ours's mechanism. 
\end{enumerate*}

Third, we compare against Model Switching (MS)~\cite{model-switching}. MS switches between single models based on the measured QPS. MS is originally intended for inference on CPUs and assumes that all models can jointly fit into RAM. Model switching has no public implementation and we therefore implemented our own version. We call this version \emph{MS+} and further enhance it:
\begin{enumerate*}[label=(\roman*)]
    \item MS+ runs on GPUs.
    \item While MS disables batching, we find that this leads to poor performance in MS+. Since MS+ is implemented on top of Clipper, we implement Clipper's batching mechanism for MS+.
    \item In our benchmarks, not all models jointly fit into VRAM. To allow MS+ to run these benchmarks, we greedily collocate as many models into the VRAM of every GPU as possible. This aims to maximize replication and throughput.
\end{enumerate*}

\subsection{End-to-end performance}
\label{sec:eval-performance}

\begin{figure}[t]
    \centering
    \vspace{-2em}
    \includegraphics[width=0.4\textwidth]{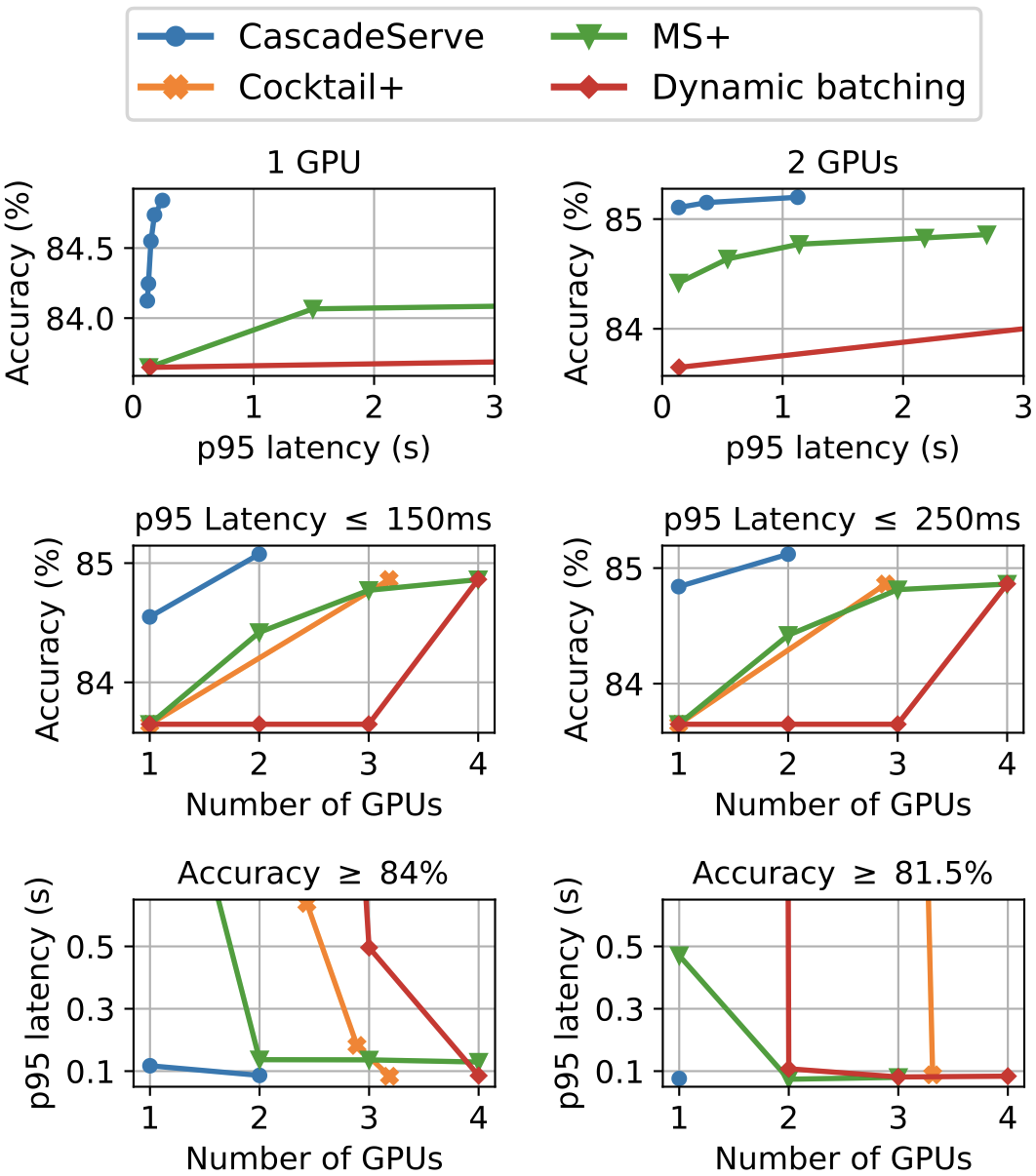}
    \caption{End-to-end performance on the BERT workload.}
    \label{fig:perf-bert}
    \vspace{-1em}
\end{figure}

We now compare \Ours's performance to the baselines described in \S\ref{sec:eval-setup}. We emulate the workload traces described in \S\ref{sec:eval-workloads} \srm{6.1 doesn't exactly describe workload traces} and iterate through the samples in the test sets of the ML benchmarks --- if we reach the end of the test set, we start from the beginning again. 
We perform an extensive grid search over the hyperparameter spaces of the baselines, and report the best hyperparameter combination found.

\begin{figure}[t]
    \centering
    \vspace{-2em}
\includegraphics[width=0.4\textwidth]{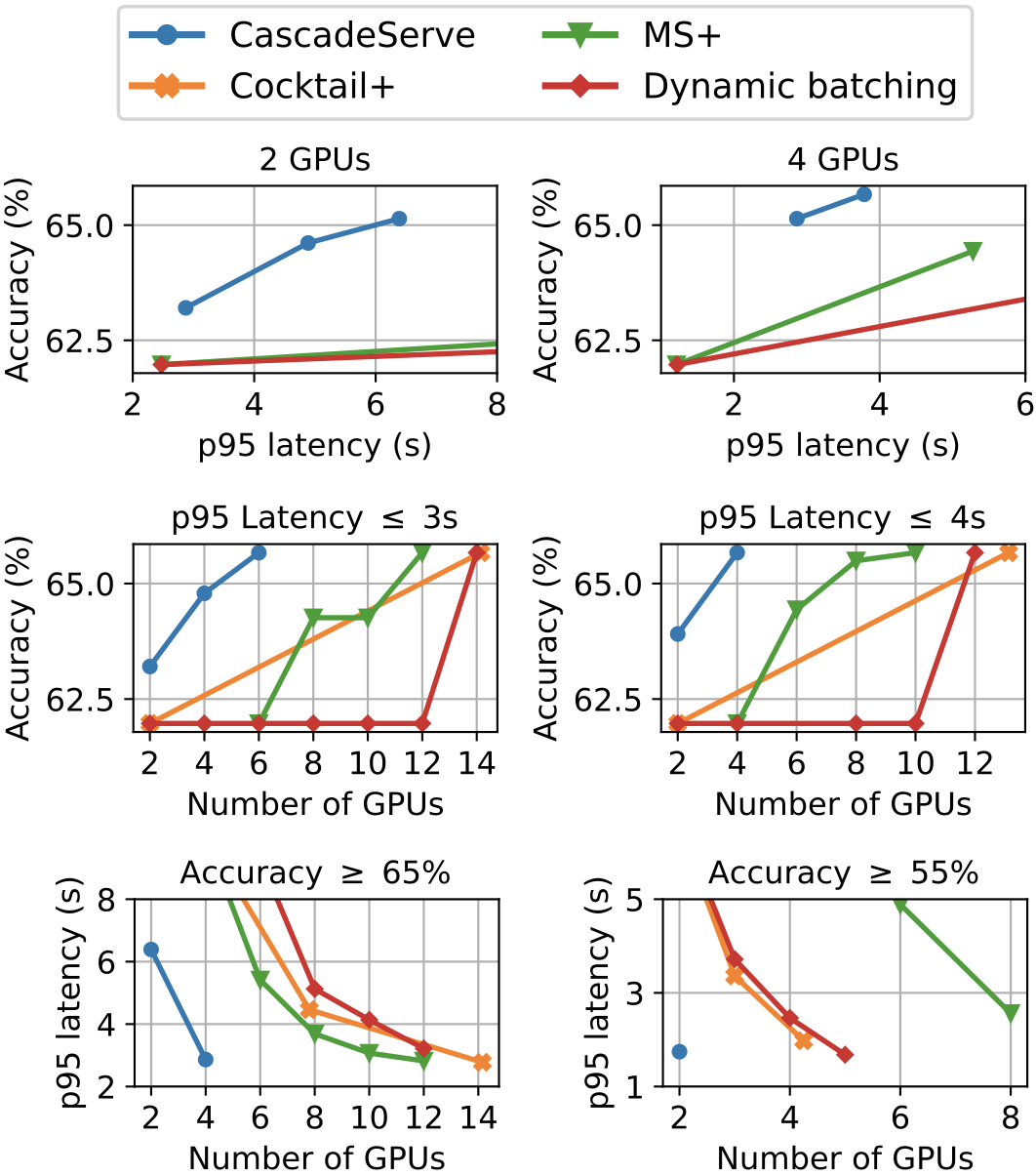}
    \caption{End-to-end performance on the Llama workload.}
    \label{fig:perf-llama}
    \vspace{-1em}
\end{figure}

The performance of each system is measured in terms of cost (number of GPUs), latency, and accuracy. In Figures~\ref{fig:perf-bert} and~\ref{fig:perf-llama}, we keep one of these three dimensions fixed and plot the trade-off between the other two. For every dimension, we use two different fixed values. 
\srm{What does the previous sentence mean -- we only switch between two possible values of each dimension? I think I'm not understanding.}
Cocktail+ doesn't support running on a fixed number of GPUs because it uses autoscaling. We therefore don't report the performance of Cocktail+ in experiments where we fix the number of GPUs.

Figures~\ref{fig:perf-bert} and ~\ref{fig:perf-llama} show that \Ours consistently outperforms the baselines for both workloads. 
\srm{We need to walk through the figures here a bit more -- use letters or numbers to denote the different subfigures and describe what each shows, including what is on the x and y axes.}
For most SLOs, \Ours can achieve peak performance using 2-3$\times$ fewer GPUs than the best baseline. For a given number of GPUs, \Ours significantly dominates the baselines at all points on the accuracy-latency curve. The dynamic batching baseline severely underutilizes its GPUs most of the time because it has to provision for workload peaks to meet latency targets. Cocktail+ improves on this by autoscaling but because of long loading times of the models, we found that good performance can only be achieved if the periodicity of autoscaling is set to a relatively large interval. As a consequence of this coarse-grained autoscaling, Cocktail+ also underutilizes its GPUs. MS+ switches between models, which may occur at a high frequency. However, MS+ faces large performance cliffs, where slight changes in the workload may require the system to switch to a model with largely inferior accuracy \srm{why?}. In comparison to these baselines, \Ours performs the best because it can make fine-grained adaptions to the workload while also avoiding large performance degradation. Furthermore, \Ours saves work by using cascades. We breakdown the impact of the two effects in \S\ref{sec:eval-ablation}.

Surprisingly, in the BERT workload, \Ours achieves a higher accuracy than the most expensive model (BERT Base). This is because the cheap models in the cascade correctly and confidently (i.e., with high certainty) classify some samples that the expensive models would have misclassified. Worth mentioning that overall, the cheap models have a lower accuracy than the expensive ones.

\begin{figure*}[t]
    \vspace{-1.2em}
    \centering
    \begin{subfigure}[b]{0.75\textwidth}
        \includegraphics[width=\linewidth]{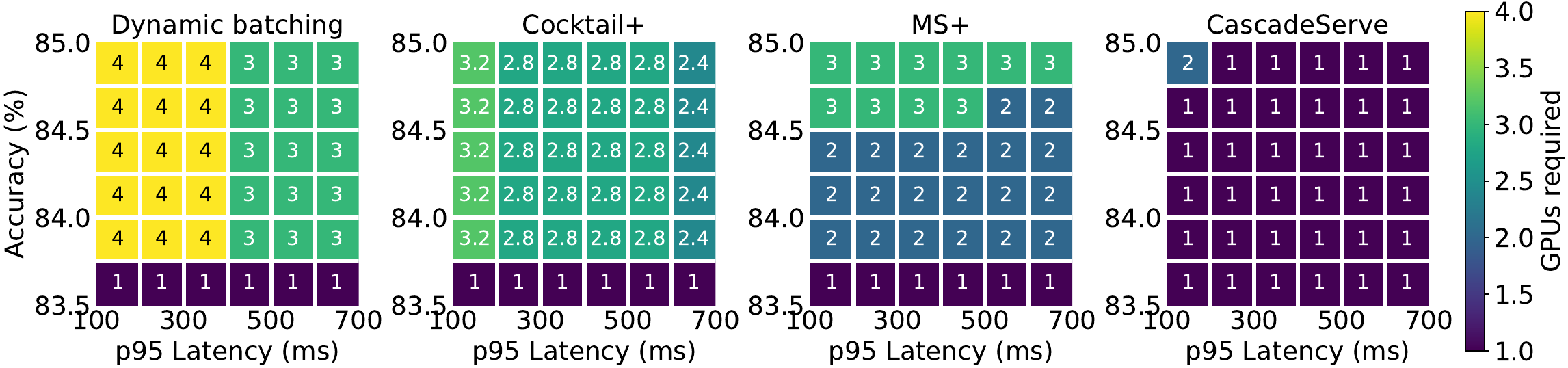}
        \caption{GPUs required to reach different accuracies and latencies for BERT.}
        \label{fig:gpus-bert}
    \end{subfigure}%
    \hspace{1em}
    \begin{subfigure}[b]{0.22\textwidth}
        \includegraphics[width=\linewidth]{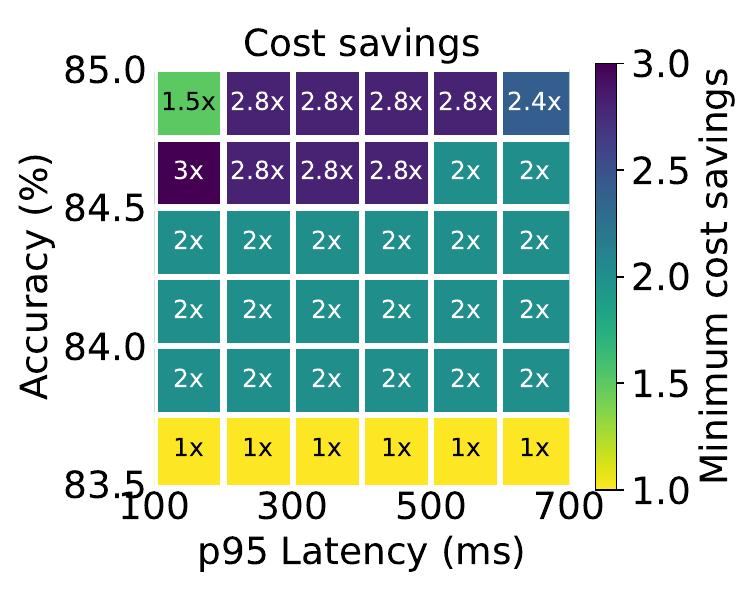}
        \caption{Cost savings for BERT}
        \label{fig:savings-bert}
    \end{subfigure}


    \begin{subfigure}[b]{0.75\textwidth}
        \vspace{0.17em}
        \includegraphics[width=\linewidth]{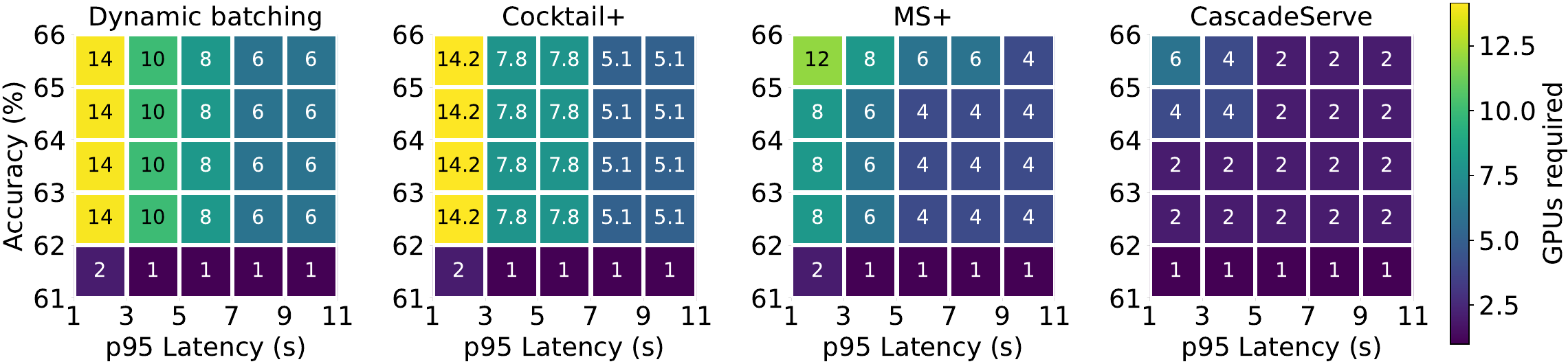}
        \caption{GPUs required to reach different accuracies and latencies for Llama.}
        \label{fig:gpus-llama}
    \end{subfigure}%
    \hspace{1.2em}
    \begin{subfigure}[b]{0.2\textwidth}
        \vspace{-0.07em}
        \includegraphics[width=\linewidth]{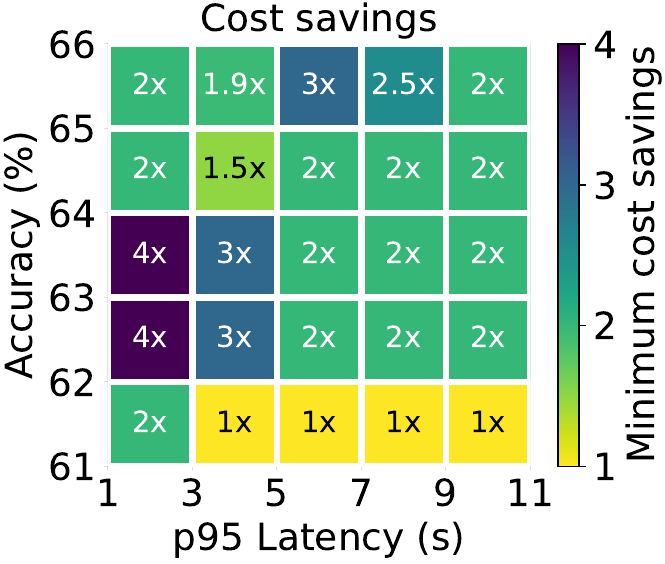}
        \caption{Cost savings for Llama}
        \label{fig:savings-llama}
    \end{subfigure}

    \caption{Required number of GPUs to reach different accuracies and latencies. The cost savings compare \Ours to the cheapest baseline at each cell.}
    \label{fig:workloads}
\end{figure*}

Figures~\ref{fig:gpus-bert} and~\ref{fig:gpus-llama} show the minimum number of GPUs that are required to achieve different points in the latency-accuracy space. Figures~\ref{fig:savings-bert} and~\ref{fig:savings-llama} derive from these results and show the cost savings that \Ours achieves when compared to the cheapest baseline at each cell. The figures show that \Ours can achieve most points in the accuracy-latency space using 2-3$\times$ fewer GPUs.
For low accuracy regimes, the serving task becomes trivial, which is why the systems don't show great performance differences on the given QPS trace. 

\subsection{Performance degradation}
\label{sec:eval-adaptivity}

When faced with high QPS, \Ours degrades accuracy and latency based on the SLO. 
Real-world workload patterns contain frequent and high variations in the query load.
To examine the behaviour of \Ours and the baselines when faced with with spiky request patterns, we now examine their performance on a simplified trace.

\begin{figure}[t]
\vspace{-0.5em}
    \centering
\includegraphics[width=0.48\textwidth]{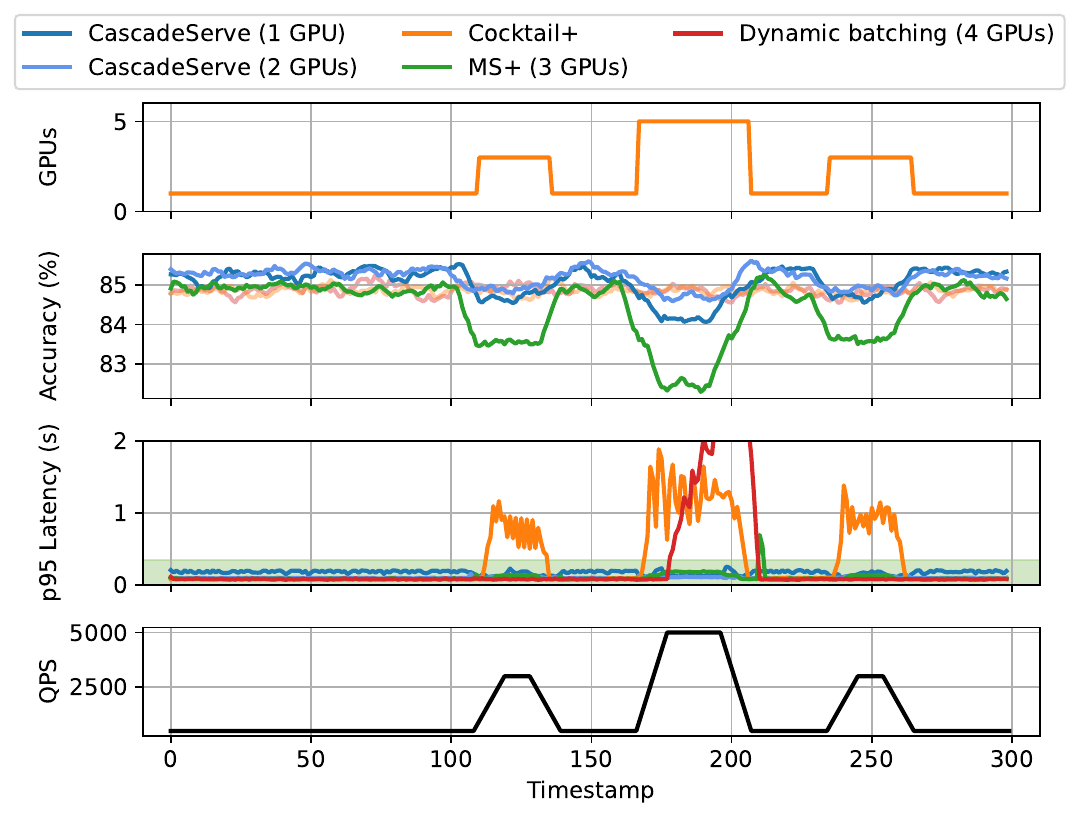}
    \caption{BERT performance degradation with latency SLO.}
    \label{fig:adapt-bert}
\vspace{-1em}
\end{figure}

In Figure~\ref{fig:adapt-bert}, we run the BERT workload with an SLO on p95 latency of at most $400ms$. We provision MS+ with three GPUs and DynBa with four GPUs. Figure~\ref{fig:adapt-bert} shows two runs where \Ours is provisioned with one and two GPUs. In the following, we focus on analyzing the run where it is provisioned with one GPU. Cocktail+ autoscales the GPUs, and we trigger replanning at QPS changes or whenever latency stabilizes again after latency spikes. We provide Cocktail+ with the ground truth QPS that will be issued. Accuracies and latencies are measured over sliding windows. The SLO is shaded in green.

Figure~\ref{fig:adapt-bert} shows how DynBa's large provisioning allows it to fulfill the latency SLO when the QPS is relatively low. However, even four GPUs are insufficient for DynBa to achieve the required throughput for the large peak. During the peak, DynBa's p95 latency shoots up to almost $4s$. Cocktail+ automatically scales its provisioning. However, loading a model into VRAM and warming it up takes several seconds. This leaves Cocktail+ under-provisioned until the resources become available and throughput can be met. Cocktail+'s latency, therefore, shoots up at every QPS peak and stays stable at high levels as soon as throughput is met again. Alternatively, MS+ will choose cheaper models for these peaks to maintain the latency while significantly decreasing the accuracy. This allows MS+ to always fulfill the latency SLO. However, even though MS+ is provisioned with three times more GPUs than \Ours, its accuracy is worse than \Ours's throughout the entire run. Finally, \Ours switches between model cascades and can fulfill the latency SLO with a minor accuracy degradation while using significantly fewer GPUs than any of the baselines.

\begin{figure}
\vspace{-0.5em}
    \centering
\includegraphics[width=0.48\textwidth]{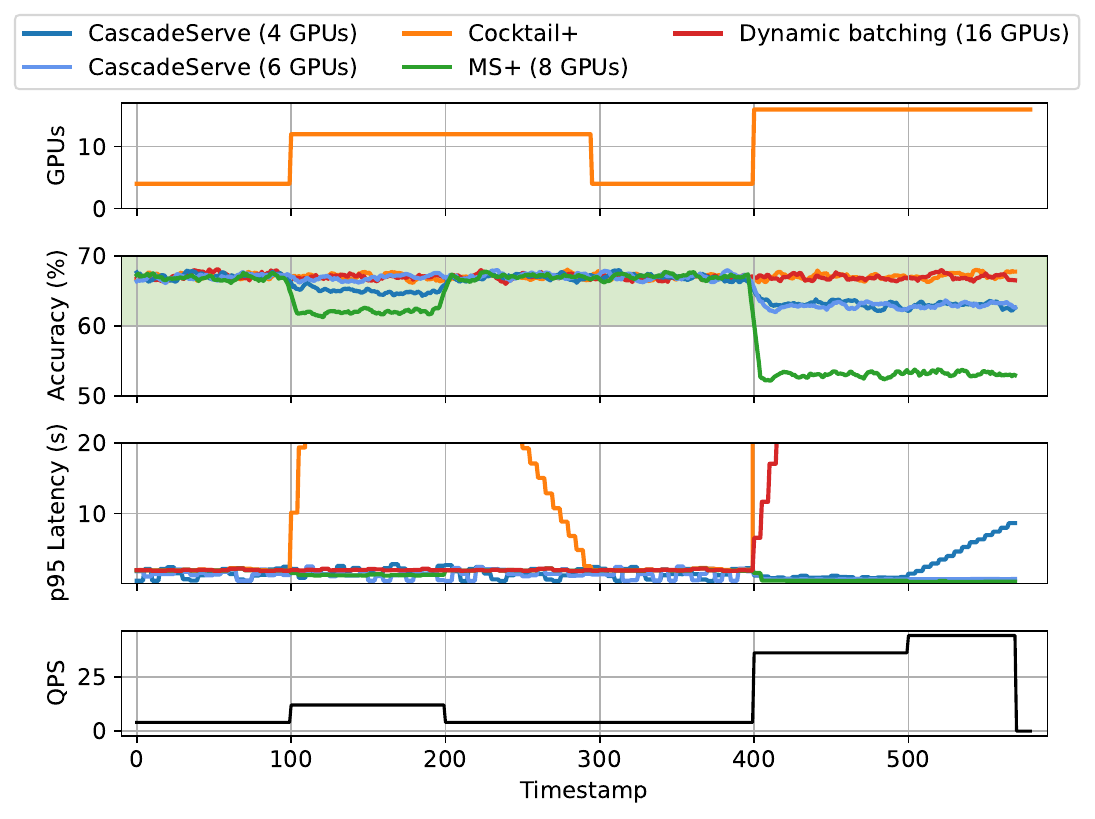}
    \caption{Llama performance degradation with accuracy SLO.}
    \label{fig:adapt-llama}
    \vspace{-1em}
\end{figure}

Figure~\ref{fig:adapt-llama} runs the Llama workload with an accuracy SLO of 60\%. Accuracies and latencies are measured over sliding windows and the SLO is shaded in green. MS+ is provisoined with eight GPUs, and DynBa with 16. We show two runs for \Ours, one with four GPUs and one with six. In the following, we focus on the run with four GPUs as it shows more clearly how \Ours degrades performance. Cocktail+ autoscales and we trigger autoscaling at QPS changes or whenever latency stabilizes again. We provide Cocktail+ with the groundtruth future QPS.

Figure~\ref{fig:adapt-llama} shows that DynBa has a static accuracy which fulfills the SLO. However, even though DynBa uses 16 GPUs, it cannot handle the second QPS spike and its latency dramatically increases. Similarly, Cocktail+'s accuracy always stays within the SLO but the large loading times of Llama models don't allow it to adapt well to the workload peaks. Instead, every QPS peak leads to a large increase in latency. MS+ achieves a low latency throughout the run at the cost of accuracy. For the larger QPS peak, MS+ violates the accuracy SLO. \Ours also degrades accuracy but only if this doesn't violate the SLO. 
After $500s$, we can see how \Ours chooses to sacrifice latency instead of accuracy in order to fulfill the SLO. Furthermore, even though MS+ is provisioned with twice as many GPUs, \Ours degrades accuracy significantly less than MS+.

\subsection{Gear plan optimizer}
\label{sec:eval-planner}

\begin{figure}
    \centering
    \vspace{-0.9em}
    \includegraphics[width=0.42\textwidth]{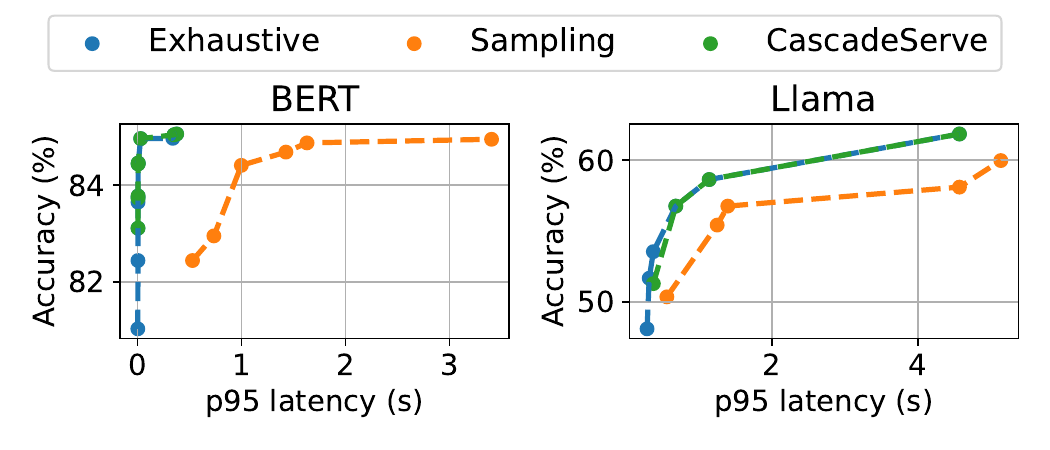}
    \caption{Quality of plans found by gear planner.}
    \label{fig:planning-quality}
\end{figure}

\begin{figure}
    \centering
    \includegraphics[width=0.47\textwidth]{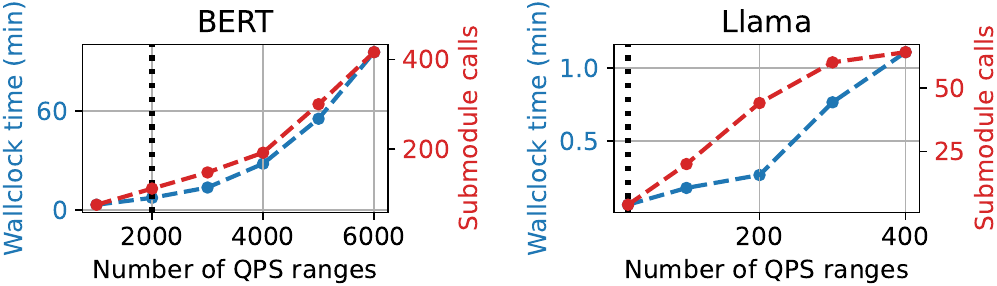}
    \caption{Cost of offline planning.}
    \label{fig:planning-time}
    \vspace{-0.7em}
\end{figure}

We now evaluate the gear plan optimizer in terms of the quality of its plans and  planning time. We evaluate all plans with respect to the metrics returned by the simulator (Appendix \ref{sec:simulator}) since the planner also operates with respect to these metrics when deployed. 

Figure~\ref{fig:planning-quality} compares the plans found by \Ours's gear planner to exhaustive search and a random sampling baseline. To make an exhaustive search feasible, we only consider a highly constrained search space. Specifically, we assume that all models can jointly be placed onto each GPU (maximum replication), which is possible since we only evaluate the gear plans on a simulator. We further set all batch sizes to be 1. Finally, to shorten the simulation time, we use a short, 8-second workload trace that is inspired by the traces in Section~\ref{sec:eval-workloads}. All methods operate on the same constrained search space. The exhaustive baseline tries all possible assignments of cascades to QPS ranges. The random sampling baseline randomly samples a cascade for each QPS range. The budget for random sampling is set to be 2$\times$ the runtime of \Ours's planner.

\Ours's planner runs for 0.1s (Llama) and 1s (BERT). The discrepancy in runtime mainly comes from the BERT workload using higher QPS, which causes higher runtimes of the simulator since more work needs to be simulated. The exhaustive search runs for 9min (Llama) and 16min (BERT). Compared to the exhaustive search, \Ours's planner misses some plans along the latency-accuracy Pareto frontier but closely approximates the Pareto-frontier. Compared to the random sampling baseline, \Ours finds significantly better plans.

Figure~\ref{fig:planning-time} evaluates the runtime of the gear planner. Here we use the full search spaces of the Llama and BERT workloads. 
The planning time is most sensitive to the number of QPS ranges $n_{ranges}$ that the plan should distinguish among (\S\ref{sec:opt-qps}). We show the total number of submodule calls and the wallclock runtime for both workloads and denote the $n_{ranges}$ used in our experiment through the vertical line. We can see that for realistic values, the offline planning time is reasonable. For example, for our experiments, each planning completed within a few minutes. Even if the user chooses to use a larger number for $n_{ranges}$, the planning time is reasonable, given that this is a one-off, offline procedure.

\subsection{Ablation study}
\label{sec:eval-ablation}

\begin{figure}
    \vspace{-0.9em}
    \centering
    \includegraphics[width=0.42\textwidth]{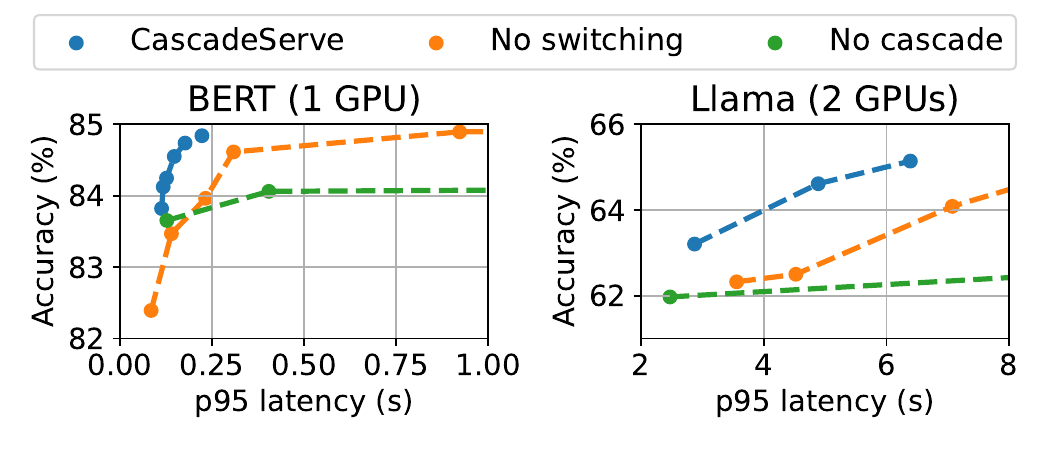}
    \caption{Performance when disabling optimizations.}
    \label{fig:ablation}
    \vspace{-1em}
\end{figure}

Finally, we examine what the drivers of \Ours's performance are in Figure {fig:ablation}. \emph{No Switching} uses a static cascade to serve inferences and \emph{No cascade} switches between single models. The performance of No cascade and MS+ might differ because they use different batching mechanisms. We run the workloads as described in Section~\ref{sec:eval-workloads}. 
Figure~\ref{fig:ablation} shows that both optimizations significantly contribute to the final performance.

\section{Conclusion}

Efficient inference serving in ML is complicated by two challenges. First, ML models incur high computational cost, and, second, the arrival rates of many practical applications have high and unpredictable variations.
In this paper we showed that model cascades are able tackle both of these challenges, as they
save work while maintaining accuracy, and
expose a high-resolution trade-off between work and accuracy.

We described \Ours, which addresses several challenges of model serving with cascades, including workload adaption, model replication onto hardware, and request batching.
\Ours operates in an offline and online phase. In the offline phase, the system pre-computes a \emph{gear plan} that specifies how to serve inferences online.
We find that \Ours saves $2-3\times$ in cost across a wide spectrum of the
latency-accuracy space when compared to state-of-the-art
baselines on different workloads.

{\footnotesize \bibliographystyle{acm}
\bibliography{references}}

\begin{thebibliography}{10}

\bibitem{sagemaker}
{AWS SageMAker}.
\newblock \url{https://aws.amazon.com/sagemaker/} (Accessed on 24 Mar 2024).

\bibitem{azureml}
{AzureML model serving}.
\newblock \url{https://learn.microsoft.com/en-us/azure/machine-learning/tutorial-deploy-model?view=azureml-api-2} (Accessed on 23 Jan 2024).

\bibitem{databricks-multimodel}
{Databricks Model Serving}.
\newblock \url{https://docs.databricks.com/en/machine-learning/model-serving/index.html} (Accessed on 21 Jan 2024).

\bibitem{databricks}
{Databricks Model Serving}.
\newblock \url{https://docs.databricks.com/en/machine-learning/model-serving/index.html} (Accessed on 24 Mar 2024).

\bibitem{triton}
{NVIDIA Triton Inference Server}.
\newblock \url{https://docs.nvidia.com/deeplearning/triton-inference-server/user-guide/docs/index.html} (Accessed on 24 Mar 2024).

\bibitem{tfx}
{TensorFlow Serving}.
\newblock \url{www.tensorflow.org/serving} (Accessed on 24 Mar 2024).

\bibitem{torchserve}
{TorchServe}.
\newblock \url{https://pytorch.org/serve/} (Accessed on 21 Jan 2024).

\bibitem{batch}
{\sc Ali, A., Pinciroli, R., Yan, F., and Smirni, E.}
\newblock Batch: Machine learning inference serving on serverless platforms with adaptive batching.
\newblock In {\em SC20: International Conference for High Performance Computing, Networking, Storage and Analysis\/} (2020), pp.~1--15.

\bibitem{scaling-laws1}
{\sc Bahri, Y., Dyer, E., Kaplan, J., Lee, J., and Sharma, U.}
\newblock Explaining neural scaling laws, 2021.

\bibitem{zipfian1}
{\sc Breslau, L., Cao, P., Fan, L., Phillips, G., and Shenker, S.}
\newblock Web caching and zipf-like distributions: evidence and implications.
\newblock In {\em IEEE INFOCOM '99. Conference on Computer Communications. Proceedings. Eighteenth Annual Joint Conference of the IEEE Computer and Communications Societies. The Future is Now (Cat. No.99CH36320)\/} (1999), vol.~1, pp.~126--134 vol.1.

\bibitem{supernet}
{\sc Cai, H., Gan, C., Wang, T., Zhang, Z., and Han, S.}
\newblock Once-for-all: Train one network and specialize it for efficient deployment.
\newblock In {\em International Conference on Learning Representations\/} (2020).

\bibitem{frugalgpt}
{\sc Chen, L., Zaharia, M., and Zou, J.}
\newblock Frugalgpt: How to use large language models while reducing cost and improving performance, 2023.

\bibitem{frugalml}
{\sc Chen, L., Zaharia, M., and Zou, J.~Y.}
\newblock Frugalml: How to use ml prediction apis more accurately and cheaply.
\newblock In {\em Advances in Neural Information Processing Systems\/} (2020), H.~Larochelle, M.~Ranzato, R.~Hadsell, M.~Balcan, and H.~Lin, Eds., vol.~33, Curran Associates, Inc., pp.~10685--10696.

\bibitem{gpulets}
{\sc Choi, S., Lee, S., Kim, Y., Park, J., Kwon, Y., and Huh, J.}
\newblock Serving heterogeneous machine learning models on {Multi-GPU} servers with {Spatio-Temporal} sharing.
\newblock In {\em 2022 USENIX Annual Technical Conference (USENIX ATC 22)\/} (Carlsbad, CA, July 2022), USENIX Association, pp.~199--216.

\bibitem{inferline}
{\sc Crankshaw, D., Sela, G.-E., Mo, X., Zumar, C., Stoica, I., Gonzalez, J., and Tumanov, A.}
\newblock Inferline: latency-aware provisioning and scaling for prediction serving pipelines.
\newblock In {\em Proceedings of the 11th ACM Symposium on Cloud Computing\/} (New York, NY, USA, 2020), SoCC '20, Association for Computing Machinery, pp.~477--491.

\bibitem{clipper}
{\sc Crankshaw, D., Wang, X., Zhou, G., Franklin, M.~J., Gonzalez, J.~E., and Stoica, I.}
\newblock Clipper: a low-latency online prediction serving system.
\newblock In {\em Proceedings of the 14th USENIX Conference on Networked Systems Design and Implementation\/} (USA, 2017), NSDI'17, USENIX Association, pp.~613--627.

\bibitem{zipfian2}
{\sc Cunha, C., Bestavros, A., and Crovella, M.}
\newblock Characteristics of www client-based traces.
\newblock Tech. rep., USA, 1995.

\bibitem{bert}
{\sc Devlin, J., Chang, M.-W., Lee, K., and Toutanova, K.}
\newblock {BERT}: Pre-training of deep bidirectional transformers for language understanding.
\newblock In {\em Proceedings of the 2019 Conference of the North {A}merican Chapter of the Association for Computational Linguistics: Human Language Technologies, Volume 1 (Long and Short Papers)\/} (Minneapolis, Minnesota, June 2019), J.~Burstein, C.~Doran, and T.~Solorio, Eds., Association for Computational Linguistics, pp.~4171--4186.

\bibitem{gptq}
{\sc Frantar, E., Ashkboos, S., Hoefler, T., and Alistarh, D.}
\newblock {GPTQ: Accurate Post-Training Quantization for Generative Pre-trained Transformers}, 2023.

\bibitem{openllama}
{\sc Geng, X., and Liu, H.}
\newblock Openllama: An open reproduction of llama, May 2023.

\bibitem{sentiment140}
{\sc Go, A., Bhayani, R., and Huang, L.}
\newblock Twitter sentiment classification using distant supervision.
\newblock CS224N Project Report 1(2009), Stanford University, 2009.

\bibitem{clockwork}
{\sc Gujarati, A., Karimi, R., Alzayat, S., Hao, W., Kaufmann, A., Vigfusson, Y., and Mace, J.}
\newblock Serving {DNNs} like clockwork: Performance predictability from the bottom up.
\newblock In {\em 14th USENIX Symposium on Operating Systems Design and Implementation (OSDI 20)\/} (Nov. 2020), USENIX Association, pp.~443--462.

\bibitem{cocktail}
{\sc Gunasekaran, J.~R., Mishra, C.~S., Thinakaran, P., Sharma, B., Kandemir, M.~T., and Das, C.~R.}
\newblock Cocktail: A multidimensional optimization for model serving in cloud.
\newblock In {\em 19th USENIX Symposium on Networked Systems Design and Implementation (NSDI 22)\/} (Renton, WA, Apr. 2022), USENIX Association, pp.~1041--1057.

\bibitem{casc-use-llm}
{\sc Gupta, N., Narasimhan, H., Jitkrittum, W., Rawat, A.~S., Menon, A.~K., and Kumar, S.}
\newblock Language model cascades: Token-level uncertainty and beyond.
\newblock In {\em The Twelfth International Conference on Learning Representations\/} (2024).

\bibitem{reef}
{\sc Han, M., Zhang, H., Chen, R., and Chen, H.}
\newblock Microsecond-scale preemption for concurrent {GPU-accelerated} {DNN} inferences.
\newblock In {\em 16th USENIX Symposium on Operating Systems Design and Implementation (OSDI 22)\/} (Carlsbad, CA, July 2022), USENIX Association, pp.~539--558.

\bibitem{holmes}
{\sc Hong, S., Xu, Y., Khare, A., Priambada, S., Maher, K., Aljiffry, A., Sun, J., and Tumanov, A.}
\newblock Holmes: Health online model ensemble serving for deep learning models in intensive care units.
\newblock In {\em Proceedings of the 26th ACM SIGKDD International Conference on Knowledge Discovery \& Data Mining\/} (New York, NY, USA, 2020), KDD '20, Association for Computing Machinery, p.~1614–1624.

\bibitem{gonzalez-multiplexing}
{\sc Jain, P., Mo, X., Jain, A., Subbaraj, H., Durrani, R.~S., Tumanov, A., Gonzalez, J., and Stoica, I.}
\newblock Dynamic space-time scheduling for gpu inference.
\newblock In {\em 32nd Conference on Neural Information Processing Systems (NeurIPS 2018)\/} (Montreal, Canada, 2018).

\bibitem{deepplan}
{\sc Jeong, J., Baek, S., and Ahn, J.}
\newblock Fast and efficient model serving using multi-gpus with direct-host-access.
\newblock In {\em Proceedings of the Eighteenth European Conference on Computer Systems\/} (New York, NY, USA, 2023), EuroSys '23, Association for Computing Machinery, pp.~249--265.

\bibitem{noscope}
{\sc Kang, D., Emmons, J., Abuzaid, F., Bailis, P., and Zaharia, M.}
\newblock Noscope: optimizing neural network queries over video at scale.
\newblock {\em Proc. VLDB Endow. 10}, 11 (aug 2017), 1586–1597.

\bibitem{scaling-laws2}
{\sc Kaplan, J., McCandlish, S., Henighan, T., Brown, T.~B., Chess, B., Child, R., Gray, S., Radford, A., Wu, J., and Amodei, D.}
\newblock Scaling laws for neural language models, 2020.

\bibitem{zipfian3}
{\sc Kavalanekar, S., Worthington, B., Zhang, Q., and Sharda, V.}
\newblock Characterization of storage workload traces from production windows servers.
\newblock In {\em 2008 IEEE International Symposium on Workload Characterization\/} (2008), pp.~119--128.

\bibitem{superserve}
{\sc Khare, A., Garg, D., Kalra, S., Grandhi, S., Stoica, I., and Tumanov, A.}
\newblock Superserve: Fine-grained inference serving for unpredictable workloads, 2023.

\bibitem{vetl}
{\sc Kossmann, F., Wu, Z., Lai, E., Tatbul, N., Cao, L., Kraska, T., and Madden, S.}
\newblock Extract-transform-load for video streams.
\newblock {\em Proc. VLDB Endow. 16}, 9 (may 2023), 2302--2315.

\bibitem{zipfian4}
{\sc Kotera, I., Egawa, R., Takizawa, H., and Kobayashi, H.}
\newblock Modeling of cache access behavior based on zipf's law.
\newblock In {\em Proceedings of the 9th Workshop on MEmory Performance: DEaling with Applications, Systems and Architecture\/} (New York, NY, USA, 2008), MEDEA '08, Association for Computing Machinery, p.~9–15.

\bibitem{willump}
{\sc Kraft, P., Kang, D., Narayanan, D., Palkar, S., Bailis, P., and Zaharia, M.}
\newblock Willump: A statistically-aware end-to-end optimizer for machine learning inference.
\newblock In {\em Proceedings of Machine Learning and Systems\/} (2020), I.~Dhillon, D.~Papailiopoulos, and V.~Sze, Eds., vol.~2, pp.~147--159.

\bibitem{producer-consumer}
{\sc Lamport, L.}
\newblock Proving the correctness of multiprocess programs.
\newblock {\em IEEE Transactions on Software Engineering SE-3}, 2 (1977), 125--143.

\bibitem{casc-search}
{\sc Lebovitz, L., Cavigelli, L., Magno, M., and Muller, L.~K.}
\newblock Efficient inference with model cascades.
\newblock {\em Transactions on Machine Learning Research\/} (2023).

\bibitem{pretzel}
{\sc Lee, Y., Scolari, A., Chun, B.-G., Santambrogio, M.~D., Weimer, M., and Interlandi, M.}
\newblock {PRETZEL}: Opening the black box of machine learning prediction serving systems.
\newblock In {\em 13th USENIX Symposium on Operating Systems Design and Implementation (OSDI 18)\/} (Carlsbad, CA, Oct. 2018), USENIX Association, pp.~611--626.

\bibitem{alpaserve}
{\sc Li, Z., Zheng, L., Zhong, Y., Liu, V., Sheng, Y., Jin, X., Huang, Y., Chen, Z., Zhang, H., Gonzalez, J.~E., and Stoica, I.}
\newblock {AlpaServe}: Statistical multiplexing with model parallelism for deep learning serving.
\newblock In {\em 17th USENIX Symposium on Operating Systems Design and Implementation (OSDI 23)\/} (Boston, MA, July 2023), USENIX Association, pp.~663--679.

\bibitem{casc-search3}
{\sc Lu, T., Wang, H., Shao, H., Gao, J., and Yao, H.}
\newblock $c^3$: Confidence calibration model cascade for inference-efficient cross-lingual natural language understanding, 2024.

\bibitem{expectation-maximization}
{\sc Moon, T.}
\newblock The expectation-maximization algorithm.
\newblock {\em IEEE Signal Processing Magazine 13}, 6 (1996), 47--60.

\bibitem{ray}
{\sc Moritz, P., Nishihara, R., Wang, S., Tumanov, A., Liaw, R., Liang, E., Elibol, M., Yang, Z., Paul, W., Jordan, M.~I., and Stoica, I.}
\newblock Ray: a distributed framework for emerging ai applications.
\newblock In {\em Proceedings of the 13th USENIX Conference on Operating Systems Design and Implementation\/} (USA, 2018), OSDI'18, USENIX Association, p.~561–577.

\bibitem{paella}
{\sc Ng, K. K.~W., Demoulin, H.~M., and Liu, V.}
\newblock Paella: Low-latency model serving with software-defined gpu scheduling.
\newblock In {\em Proceedings of the 29th Symposium on Operating Systems Principles\/} (New York, NY, USA, 2023), SOSP '23, Association for Computing Machinery, pp.~595--610.

\bibitem{casc-search2}
{\sc Nie, L., Ding, Z., Hu, E., Jermaine, C., and Chaudhuri, S.}
\newblock Online cascade learning for efficient inference over streams, 2024.

\bibitem{pytorch}
{\sc Paszke, A., Gross, S., Massa, F., Lerer, A., Bradbury, J., Chanan, G., Killeen, T., Lin, Z., Gimelshein, N., Antiga, L., Desmaison, A., K\"{o}pf, A., Yang, E., DeVito, Z., Raison, M., Tejani, A., Chilamkurthy, S., Steiner, B., Fang, L., Bai, J., and Chintala, S.}
\newblock {\em PyTorch: an imperative style, high-performance deep learning library}.
\newblock Curran Associates Inc., Red Hook, NY, USA, 2019.

\bibitem{casc-use-recommend}
{\sc Rebelo, M.~{\^A}., Coelho, D., Pereira, I., and Fernandes, F.}
\newblock A new cascade-hybrid recommender system approach for the retail market.
\newblock In {\em Innovations in Bio-Inspired Computing and Applications\/} (Cham, 2022), A.~Abraham, A.~M. Madureira, A.~Kaklauskas, N.~Gandhi, A.~Bajaj, A.~K. Muda, D.~Kriksciuniene, and J.~C. Ferreira, Eds., Springer International Publishing, pp.~371--380.

\bibitem{yolo}
{\sc Redmon, J., Divvala, S., Girshick, R., and Farhadi, A.}
\newblock You only look once: Unified, real-time object detection.
\newblock In {\em 2016 IEEE Conference on Computer Vision and Pattern Recognition (CVPR)\/} (Los Alamitos, CA, USA, jun 2016), IEEE Computer Society, pp.~779--788.

\bibitem{infaas}
{\sc Romero, F., Li, Q., Yadwadkar, N.~J., and Kozyrakis, C.}
\newblock {INFaaS}: Automated model-less inference serving.
\newblock In {\em 2021 USENIX Annual Technical Conference (USENIX ATC 21)\/} (July 2021), USENIX Association, pp.~397--411.

\bibitem{monitor-online}
{\sc Schröder, T., and Schulz, M.}
\newblock Monitoring machine learning models: a categorization of challenges and methods.
\newblock {\em Data Science and Management 5}, 3 (2022), 105--116.

\bibitem{azure-trace}
{\sc Shahrad, M., Fonseca, R., Goiri, I., Chaudhry, G., Batum, P., Cooke, J., Laureano, E., Tresness, C., Russinovich, M., and Bianchini, R.}
\newblock Serverless in the wild: Characterizing and optimizing the serverless workload at a large cloud provider.
\newblock In {\em 2020 USENIX Annual Technical Conference (USENIX ATC 20)\/} (July 2020), USENIX Association, pp.~205--218.

\bibitem{nexus}
{\sc Shen, H., Chen, L., Jin, Y., Zhao, L., Kong, B., Philipose, M., Krishnamurthy, A., and Sundaram, R.}
\newblock Nexus: a gpu cluster engine for accelerating dnn-based video analysis.
\newblock In {\em Proceedings of the 27th ACM Symposium on Operating Systems Principles\/} (New York, NY, USA, 2019), SOSP '19, Association for Computing Machinery, pp.~322--337.

\bibitem{weg}
{\sc Shen, H., Han, S., Philipose, M., and Krishnamurthy, A.}
\newblock Fast video classification via adaptive cascading of deep models.
\newblock In {\em 2017 IEEE Conference on Computer Vision and Pattern Recognition (CVPR)\/} (2017), pp.~2197--2205.

\bibitem{orion}
{\sc Strati, F., Ma, X., and Klimovic, A.}
\newblock Orion: Interference-aware, fine-grained gpu sharing for ml applications.
\newblock In {\em Nineteenth European Conference on Computer Systems (EuroSys '24)\/} (Athens, Greece, April 22--25 2024), ACM, pp.~1--18.

\bibitem{llama-2}
{\sc Touvron, H., Martin, L., Stone, K., Albert, P., Almahairi, A., Babaei, Y., Bashlykov, N., Batra, S., Bhargava, P., Bhosale, S., Bikel, D., Blecher, L., Ferrer, C.~C., Chen, M., Cucurull, G., Esiobu, D., Fernandes, J., Fu, J., Fu, W., Fuller, B., Gao, C., Goswami, V., Goyal, N., Hartshorn, A., Hosseini, S., Hou, R., Inan, H., Kardas, M., Kerkez, V., Khabsa, M., Kloumann, I., Korenev, A., Koura, P.~S., Lachaux, M.-A., Lavril, T., Lee, J., Liskovich, D., Lu, Y., Mao, Y., Martinet, X., Mihaylov, T., Mishra, P., Molybog, I., Nie, Y., Poulton, A., Reizenstein, J., Rungta, R., Saladi, K., Schelten, A., Silva, R., Smith, E.~M., Subramanian, R., Tan, X.~E., Tang, B., Taylor, R., Williams, A., Kuan, J.~X., Xu, P., Yan, Z., Zarov, I., Zhang, Y., Fan, A., Kambadur, M., Narang, S., Rodriguez, A., Stojnic, R., Edunov, S., and Scialom, T.}
\newblock Llama 2: Open foundation and fine-tuned chat models, 2023.

\bibitem{exllama}
{\sc turboderp}.
\newblock Exllama github repository.
\newblock \url{https://github.com/turboderp/exllama}, 2024.

\bibitem{bert-variants}
{\sc Turc, I., Chang, M.-W., Lee, K., and Toutanova, K.}
\newblock Well-read students learn better: On the importance of pre-training compact models, 2019.

\bibitem{first-cascade}
{\sc Viola, P., and Jones, M.}
\newblock Rapid object detection using a boosted cascade of simple features.
\newblock In {\em Proceedings of the 2001 IEEE Computer Society Conference on Computer Vision and Pattern Recognition. CVPR 2001\/} (2001), vol.~1, pp.~I--I.

\bibitem{rafiki}
{\sc Wang, W., Gao, J., Zhang, M., Wang, S., Chen, G., Ng, T.~K., Ooi, B.~C., Shao, J., and Reyad, M.}
\newblock Rafiki: machine learning as an analytics service system.
\newblock {\em Proc. VLDB Endow. 12}, 2 (oct 2018), 128--140.

\bibitem{idk-cascades}
{\sc Wang, X., Luo, Y., Crankshaw, D., Tumanov, A., Yu, F., and Gonzalez, J.~E.}
\newblock Idk cascades: Fast deep learning by learning not to overthink, 2018.

\bibitem{huggingface}
{\sc Wolf, T., Debut, L., Sanh, V., Chaumond, J., Delangue, C., Moi, A., Cistac, P., Rault, T., Louf, R., Funtowicz, M., Davison, J., Shleifer, S., von Platen, P., Ma, C., Jernite, Y., Plu, J., Xu, C., Scao, T.~L., Gugger, S., Drame, M., Lhoest, Q., and Rush, A.~M.}
\newblock Huggingface's transformers: State-of-the-art natural language processing, 2020.

\bibitem{meta-scale}
{\sc Wu, C.-J., Raghavendra, R., Gupta, U., Acun, B., Ardalani, N., Maeng, K., Chang, G., Aga, F., Huang, J., Bai, C., Gschwind, M., Gupta, A., Ott, M., Melnikov, A., Candido, S., Brooks, D., Chauhan, G., Lee, B., Lee, H.-H., Akyildiz, B., Balandat, M., Spisak, J., Jain, R., Rabbat, M., and Hazelwood, K.}
\newblock Sustainable ai: Environmental implications, challenges and opportunities.
\newblock In {\em Proceedings of Machine Learning and Systems\/} (2022), D.~Marculescu, Y.~Chi, and C.~Wu, Eds., vol.~4, pp.~795--813.

\bibitem{irina}
{\sc Wu, X., Xu, H., and Wang, Y.}
\newblock Irina: Accelerating dnn inference with efficient online scheduling.
\newblock In {\em Proceedings of the 4th Asia-Pacific Workshop on Networking\/} (New York, NY, USA, 2020), APNet '20, Association for Computing Machinery, pp.~36--43.

\bibitem{stage}
{\sc Wu, Z., Marcus, R., Liu, Z., Negi, P., Nathan, V., Pfeil, P., Saxena, G., Rahman, M., Narayanaswamy, B., and Kraska, T.}
\newblock Stage: Query execution time prediction in amazon redshift, 2024.

\bibitem{twitter-acc-deg}
{\sc {X Engineering Blog}}.
\newblock Resilient ad serving at twitter-scale, March 2016.
\newblock \url{https://blog.x.com/engineering/en_us/a/2016/resilient-ad-serving-at-twitter-scale} (accessed on May 7, 2024).

\bibitem{unfoldml}
{\sc Xu, Y., Khare, A., Matlin, G., Ramadoss, M., Kamaleswaran, R., Zhang, C., and Tumanov, A.}
\newblock Unfoldml: Cost-aware and uncertainty-based dynamic 2d prediction for multi-stage classification, 2022.

\bibitem{zipfian6}
{\sc Yang, Y., and Zhu, J.}
\newblock Write skew and zipf distribution: Evidence and implications.
\newblock {\em ACM Trans. Storage 12}, 4 (jun 2016).

\bibitem{zipfian5}
{\sc Yu, H., Zheng, D., Zhao, B.~Y., and Zheng, W.}
\newblock Understanding user behavior in large-scale video-on-demand systems.
\newblock In {\em Proceedings of the 1st ACM SIGOPS/EuroSys European Conference on Computer Systems 2006\/} (New York, NY, USA, 2006), EuroSys '06, Association for Computing Machinery, p.~333–344.

\bibitem{salus}
{\sc Yu, P., and Chowdhury, M.}
\newblock Salus: Fine-grained gpu sharing primitives for deep learning applications, 2019.

\bibitem{hellaswag}
{\sc Zellers, R., Holtzman, A., Bisk, Y., Farhadi, A., and Choi, Y.}
\newblock {H}ella{S}wag: Can a machine really finish your sentence?
\newblock In {\em Proceedings of the 57th Annual Meeting of the Association for Computational Linguistics\/} (Florence, Italy, July 2019), A.~Korhonen, D.~Traum, and L.~M{\`a}rquez, Eds., Association for Computational Linguistics, pp.~4791--4800.

\bibitem{mark}
{\sc Zhang, C., Yu, M., Wang, W., and Yan, F.}
\newblock {MArk}: Exploiting cloud services for {Cost-Effective}, {SLO-Aware} machine learning inference serving.
\newblock In {\em 2019 USENIX Annual Technical Conference (USENIX ATC 19)\/} (Renton, WA, July 2019), USENIX Association, pp.~1049--1062.

\bibitem{shepherd}
{\sc Zhang, H., Tang, Y., Khandelwal, A., and Stoica, I.}
\newblock {SHEPHERD}: Serving {DNNs} in the wild.
\newblock In {\em 20th USENIX Symposium on Networked Systems Design and Implementation (NSDI 23)\/} (Boston, MA, Apr. 2023), USENIX Association, pp.~787--808.

\bibitem{model-switching}
{\sc Zhang, J., Elnikety, S., Zarar, S., Gupta, A., and Garg, S.}
\newblock {Model-Switching}: Dealing with fluctuating workloads in {Machine-Learning-as-a-Service} systems.
\newblock In {\em 12th USENIX Workshop on Hot Topics in Cloud Computing (HotCloud 20)\/} (July 2020), USENIX Association.

\end{thebibliography}


\appendix
\section{Gear planner convergence proof}
\label{sec:convergence}

In this section we prove that the gear plan optimization algorithm described in Section~\ref{sec:offline} always returns. The argument is analogous to convergence proofs for Expectation-Maximization algorithms~\cite{expectation-maximization}.
We distinguish between two cases.

\paragraph{Case 1: The algorithm never finds a feasible plan.} The planner will keep iterating through the submodules and the submodules will always throw an error whenever they are called. This causes the workload adaption submodule (\verb|SP2|) to downgrade a cascade in each iteration (i.e., if the SLO is on accuracy, it will use more accurate cascade, and if the SLO is on latency, it will use a cascade with more throughput). This process can only be repeated until there is no QPS whose cascade can further be downgraded. At that point, the module will also throw an error and the cascade search submodule (\verb|SP1|) will raise an error to the user. The algorithm has terminated.

\paragraph{Case 2: The algorithm finds a feasible plan at some point.}

Once the algorithm finds a feasible plan, it will never produce an infeasible plan again. This is insured by each submodule, which only updates the gear plan if this update improves the plan while making sure it remains feasible. This also means that in each iteration of the algorithm, the gear plan only becomes better (in terms of the non-SLO metric) or stays the same. If the plan stays the same for one entire iteration through all submodules, the algorithm terminates. It is guaranteed that such an iteration eventually occurs, since the gear plan cannot be improved indefinitely: The latency cannot go below $0ms$ and the prediction quality cannot go beyond perfect predictions (e.g. 100\% accuracy).

\section{Certainty estmation}
\label{sec:certainty}

The method for certainty estimation is not fundamental to \Ours and can be replaced with other methods, e.g., methods proposed in other works~\cite{idk-cascades}. For our experiments, we found that the following method works well and used it in all of the experiments.

Many ML tasks involve assigning scores to options. For example, recommender systems assign scores to objects, classifiers assign scores to classes, text generation models assign scores to tokens, etc. To estimate certainty, we simply take the score of the highest scoring entity and subtract the score of the second-highest scoring entity. High differences mean that the model is clear about what the highest scoring entity is --- this corresponds to a high prediction certainty. Low differences mean that it is contested which entity scores highest (e.g. which class the input belongs to) --- this corresponds to low prediction certainties. 

Equation~\ref{eq:cert} denotes the certainty of model $model$ on input $x$. $model(x)^{(1)}$ is the entity with the highest score and $model(x)^{(2)}$ the entity with the second highest score (e.g., the most likely class and the second most likely class according to the model).

\begin{equation}
\label{eq:cert}
    cert(model, x) = model(x)^{(1)} - model(x)^{(2)}
\end{equation}

\section{Simulator}
\label{sec:simulator}

\Ours uses a simulator to evaluate cascades and gear plans during the offline phase. It is a known results that model serving systems can accurately be simulated~\cite{alpaserve, clockwork}. In \Ours, this simulator is used to drastically improve the runtime of gear plan generation.

\subsection{Latency simulation}

\begin{figure}
    \centering
    \includegraphics[width=0.48\textwidth]{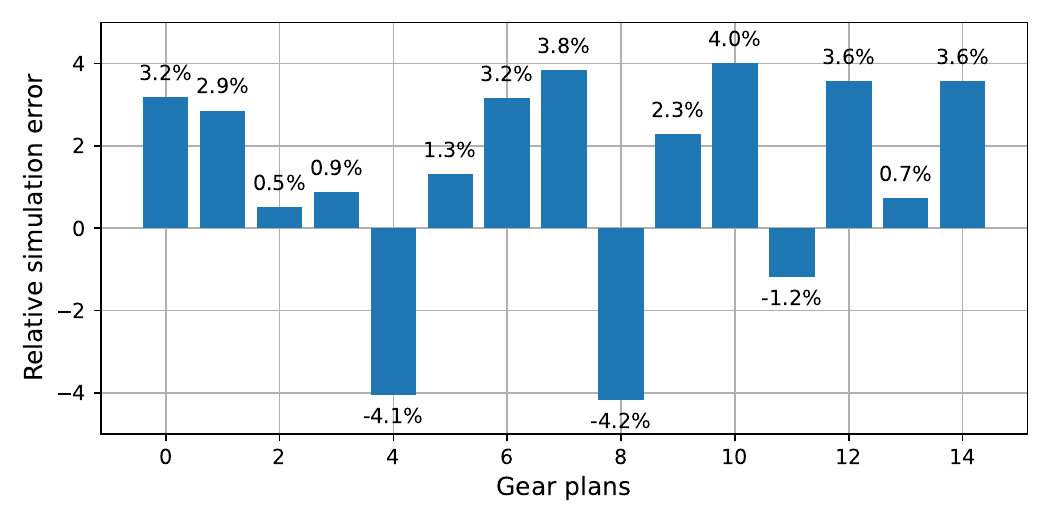}
    \caption{Relative error in p95 latency simulation for 15 gear plans.. Positive errors mean the latency was over-estimated, negative errors mean the latency was under-estimated.}
    \label{fig:sim-eval}
\end{figure}

We describe how an individual cascade is simulated for a given QPS. Simulating different cascades at different QPSes follows trivially by continuing the simulation while not resetting the simulation state from the previous QPS.

In a first step, \Ours profiles all models that the user registered with different batch sizes. \Ours further lets every model predict on the validation data used in the simulation, and records its prediction and certainty. \srm{Explain previously that the simulation takes in a set of validation data.}

In a nutshell, \Ours's simulator closely mirrors what the system does when actually running. The gear is periodically switched in the simulation, as it is in real runs. As queries are issued, samples are put in the queue of the first model of the cascade specified in the current gear. The queue lengths of all models are polled with every query being added. Once a queue length reaches the minimum batch size defined in the gear, its inference is triggered, pending on the simulated GPU becoming available. Once the GPU is available, inference is simulated on all samples in the queue by reading out the profiled runtime of the model for the given queue size. During that runtime, the GPU is blocked and other inferences may not be scheduled on it. 

For runtime estimation, the  simulator simply uses the certainties and runtimes of the samples in the validation set and cycles through them. The simulator cascades a subset of the samples in a batch based on the pre-recorded prediction certainties and the certainty threshold set in the gear. 

We evaluate the simulator in Figure~\ref{fig:sim-eval}. Each bar in the chart corresponds to a gear plan, whose p95 latency was simulated and then compared to the runtime of an actual run. Figure~\ref{fig:sim-eval} reports the percentage difference between the two latencies. Figure~\ref{fig:sim-eval} shows 15 different gear plans, on three different workload traces and for two different models (BERT and Llama).

\subsection{Accuracy simulation}

We now describe how \Ours simulates accuracy. \Ours uses these accuracy estimates to make decisions on which models and which cascades to use for online serving. Deciding on which models to use based on the model's offline performance on validation data is a common practice, which is also employed by other serving systems~\cite{infaas, cocktail}. Nevertheless, the offline accuracy measurement may differ from what users experience online, which is a related to monitoring ML services, a problem addressed in literature orthogonal to this work~\cite{monitor-online}. \Ours can only estimate the online accuracy when making strong assumptions about the workload.

The accuracy of a cascade is determined by predicting all validation data using the cascade and computing the accuracy using the user-provided, ground truth labels. The accuracy of a gear plan is determined as the time-weighted average of all cascades being used in the gear plan. Most time intervals in real-world workloads typically comprise of a low QPS, with occasional workload spikes~\cite{azure-trace}. Prior work in related areas has found that the load distributions of real-world workloads typically follow a Zipf-like distribution~\cite{zipfian1}. This has inspired works of many domains to model their workload as a Zipfian~\cite{zipfian2, zipfian3, zipfian4, zipfian5, zipfian6}. In \Ours, the simulator also assumes per default that the QPS distribution is Zipfian --- that is, when the gear planner needs to choose a QPS interval for downgrading, it will assess the cascades' impact on the overall accuracy while considering that low-QPS regimes occur more frequently than high-QPS regimes. During online serving, \Ours measures the QPS in any case as an artifact of gear switching. If the system finds that the actual QPS distribution strongly deviates from the expected one, it will notify the user. The user may then choose whether they want to keep the current gear plan or trigger re-planning with a more accurate qPS distribution (e.g., the one that \Ours measured).

If the user sets an SLO on latency, the gear planner will ensure that the latency target is obeyed for any QPS. This means that the Zipfian assumption does not have an effect on SLO attainment but might have an effect on the optimality of the plan if the online distribution strongly deviates from a Zipfian. In this case the gear planner might have deprioritized the wrong cascades for downgrading.

If the user sets an SLO on accuracy, the gear planner will attempt to meet this SLO for the assumed Zipfian distribution. Specifically the gear planner will assign accurate cascades to low-QPS regimes and assume that these cascades are used for longer time periods than the less accurate cascades assigned to high-QPS regimes. To avoid SLO violations online, \Ours scales the maps the measured QPS distribution onto the distribution used in the gear plan --- this is realized by scaling each measured QPS value by a linear factor, such that measured QPS distribution after scaling matches the one of the gear plan. Like for the latency SLO, the Zipfian assumtion does therefore not lead to SLO violations but may lead to suboptimal gear plans if the assumed distribution strongly deviates from the one found online. In the most extreme scenario, the gear plan might be scaled such that it degenerates to always using the most expensive cascade.

\end{document}